\documentclass{article}

\usepackage{PRIMEarxiv}
\usepackage{graphicx}
\usepackage[outdir=./]{epstopdf}
\usepackage{authblk}
\usepackage[utf8]{inputenc} 
\usepackage[T1]{fontenc}    
\usepackage{hyperref}       
\usepackage{url}            
\usepackage{booktabs}       
\usepackage{amsfonts}       
 \usepackage{amsmath}       
\usepackage{nicefrac}       
\usepackage{microtype}      
\usepackage{lipsum}
\usepackage{fancyhdr}       
\usepackage{graphicx}       
\graphicspath{{media/}}     

\title{PtyLab.m/py/jl: a cross-platform, open-source inverse modeling toolbox for conventional and Fourier ptychography}

\author[1,2,3,4,*]{Lars Loetgering*}
\author[5]{Mengqi Du}
\author[4]{Dirk Boonzajer Flaes}
\author[6]{Tomas Aidukas}
\author[4]{Felix Wechsler}
\author[1,2,3,8]{Daniel S. Penagos Molina}
\author[$\dagger$]{Max Rose}
\author[5,7]{Antonios Pelekanidis}
\author[1,2,3]{Wilhelm Eschen}
\author[8]{Jürgen Hess}
\author[9]{Thomas Wilhein}
\author[4,10]{Rainer Heintzmann}
\author[1,2,3,8]{Jan Rothhardt}
\author[5,7]{Stefan Witte}

\affil[1]{Institute of Applied Physics and Abbe Center of Photonics, Friedrich-Schiller-University Jena, Germany}
\affil[2]{Helmholtz-Institute Jena, Germany}
\affil[3]{GSI Helmholtzzentrum für Schwerionenforschung, Darmstadt, Germany}
\affil[4]{Leibniz Institute of Photonic Technology, Jena, Germany}
\affil[5]{ARCNL, Amsterdam, The Netherlands}
\affil[6]{Paul Scherrer Institute, Villigen, Switzerland}
\affil[7]{Dept. of Physics and Astronomy, Vrije Universiteit Amsterdam, The Netherlands}
\affil[8]{Fraunhofer Institute for Applied Optics and Precision Engineering, Jena, Germany}
\affil[9]{RheinAhrCampus, Remagen, Germany}
\affil[10]{Institute of Physical Chemistry and Abbe Center of Photonics, Friedrich-Schiller-University Jena, Germany}
\affil[$\dagger$]{no affiliation}
\affil[*]{lars.loetgering@fulbrightmail.org}

\begin{document}
\maketitle





\begin{abstract}
Conventional (CP) and Fourier (FP) ptychography have emerged as versatile quantitative phase imaging techniques. While the main application cases for each technique are different, namely lens-less short wavelength imaging for CP and lens-based visible light imaging for FP, both methods share a common algorithmic ground. CP and FP have in part independently evolved to include experimentally robust forward models and inversion techniques. This separation has resulted in a plethora of algorithmic extensions, some of which have not crossed the boundary from one modality to the other. Here, we present an  open source, cross-platform software, called \emph{PtyLab}, enabling both CP and FP data analysis in a unified framework. With this framework, we aim to facilitate and accelerate cross-pollination between the two techniques. Moreover, the availability in Matlab, Python, and Julia will set a low barrier to enter each field. 
\end{abstract}


\section{Introduction}

Ptychography \cite{Faulkner2004a, Rodenburg2019}  has grown into a mature technique for x-ray, extreme ultraviolet (EUV), and electron microscopy. It has revolutionized synchrotron-based x-ray microscopy, where it improves upon previously existing scanning transmission x-ray microscopy (STXM) data analysis techniques \cite{Pattee1953, Horowitz1972,Rarback1988,Jacobsen1991}. Three major benefits of ptychography over STXM are: (1) decoupling of the illumination spot size from the achievable lateral resolution, (2) quantitative amplitude and phase contrast, and (3) access to wavefront diagnostics \cite{Rodenburg2007a, Thibault2008, Thibault2009_contrast_mechanisms, Pfeiffer2018, wang2023optical}. Similar benefits have subsequently been demonstrated for scanning transmission electron microscopes (STEMs) \cite{Hue2010, Putkunz2012,Humphry2012}, where it recently produced micrographs at record-breaking resolution \cite{Jiang2018,chen2021electron}. A parallel line of development is EUV laboratory-scale microscopy, where ptychography is a promising candidate for actinic inline metrology for lithography applications \cite{Seaberg2014, Gardner2017Subwavelength, Loetgering2022Advances} and a tool for chemically-resolved microscopy \cite{Tanksalvala2021, eschen2022material}. In ptychography, a specimen is laterally scanned through a localized illumination, referred to as probe. A detector downstream of the specimen records a sequence of diffraction patterns. These observations lack phase information preventing direct inversion. Ptychography solves this problem by recording data from laterally overlapping specimen regions of interest. This acquisition scheme opens up the possibility for phase retrieval and simultaneous deconvolution of illumination and specimen information. Beyond operation with x-ray and electron radiation, ptychography has been implemented with extreme ultraviolet, visible, near-infrared, and terahertz radiation \cite{Maiden2011, Marrison2013, Du2021, Valzania2018,Loetgering2022Advances}.

Fourier ptychography \cite{Zheng2013} follows a similar operational principle as (conventional) ptychography, denoted FP and CP, respectively, throughout this paper. In FP, a specimen is illuminated from different directions, typically steered by means of an LED array, which serves as a controllable condenser. A sequence of low-resolution bright and dark field images is recorded in a lens-based microscope. Changing the illumination direction amounts to shifting the object spectrum with respect to the pupil of the optical system. If the illumination direction is changed in such a way that two recorded images share information in the Fourier domain, phase retrieval techniques can be applied to separately reconstruct the object spectrum and the pupil of the optical system. Thus FP has three attractive features: (1) The low-resolution data can be stitched together to a large synthetic numerical aperture (NA), resulting in both a large field of view and high resolution. In contrast to most wide-field systems, FP thus does not trade-off resolution and field of view. (2) after conversion to real-space, the recovered object spectrum gives quantitative amplitude and phase maps of the sample; (3) the reconstructed pupil function enables aberration diagnostics of the optical system at hand \cite{Ou2014, Konda2020, Zheng2021}. While FP has mostly found applications in the visible domain, recent implementations using infrared radiation and x-rays have been reported \cite{Sen2016Fourier,Wakonig2019Xray}. 

\subsection{Contribution}

In both CP and FP, the recorded data jointly sample real and reciprocal space \cite{Rodenburg1992,Chapman1996, li2014WDD, Horstmeyer2014,DaSilva2015}. In CP, the probe serves as a real space window that selects local spatial frequency content. In FP, the pupil selects a low-resolution real space image from a localized Fourier space bandpass. In fact, the forward models of CP and FP are mathematically equivalent and the measured data cubes may be regarded as rotations of one another in phase space \cite{Horstmeyer2014}. Although this equivalence is well-known, CP and FP have evolved into two separate communities with different algorithmic approaches and self-calibration techniques. Here, we report on a numerical data analysis toolbox, named \emph{PtyLab} (code available online \cite{PtyLab}), which places the equivalence of CP and FP at the center of its logical structure, resulting in three main contributions of this work:
\begin{enumerate}
    \item Cross-modal: \emph{PtyLab} allows to not only analyze CP and FP data, but also convert the same data set between the two domains. This flexible conversion between CP and FP leads to both physical insights as well as algorithmic cross-pollination of both domains. To our knowledge, \emph{PtyLab} is the first ptychography code designed to be cross-modal, unifying the data analysis frameworks of CP and FP.  
    \item Multi-lingual: \emph{PtyLab} is the first cross-platform ptychography code available in three programming languages, namely Matlab, Python, and Julia. Thus, it enables researchers with different programming backgrounds to communicate and exchange ideas based on a unified terminology and and code structure.
    \item Open access: \emph{PtyLab} is released together with various experimental data sets and accompanying hands-on tutorials, where the user is trained in practical data analysis. We hope that this contributes to standardized data analysis in both CP and FP. 
\end{enumerate}

In addition, \emph{PtyLab} features a variety of algorithmic extensions as compared to currently available ptychography software packages. Some of these were previously reported by us and are now provided open source. This includes axial (zPIE) \cite{Loetgering2020zPIE} as well as angle (aPIE) \cite{deBeurs2022apie} correction engines, code to analyze ptychographic optical coherence tomography (POCT) data, efficient wave propagation algorithms both valid for high NA and polychromatic radiation, and detector subsampling (sPIE) \cite{Loetgering2020zPIE,deBeurs2022apie,Du2021,Loetgering2021,Loetgering2017a}. Other novelties are reported here for the first time such as external reference wave ptychography. In addition, previously reported algorithms developed by other groups are included such as the extended ptychographic iterative engine (ePIE) \cite{Maiden2009}, multislice (e3PIE) \cite{Maiden2012multislice}, mixed states \cite{Thibault2013}, information multiplexing (PIM) \cite{Batey2014}, momentum acceleration (mPIE) \cite{Maiden2017}, Tikhonov and total variation regularization \cite{Thibault2012,Stockmar2015,Tanksalvala2021}, correlation-based lateral position correction \cite{Zhang2013}, and orthogonal probe relaxation (OPR) \cite{Odstrcil2016OPR}. In writing this manuscript, we pursued the goal of providing a concise overview of the various engines available to date in ptychography. 

\subsection{Related work}

Most CP packages reported to date have focused on high performance computing, which is key for day-to-day user operation at large scale facilities, where beamtime is scarce and experimental feedback is needed quickly \cite{nashed2014parallel, marchesini2016sharp, Enders2016, Odstrcil2018,Dong2018,wakonig2020ptychoshelves, Favre-Nicolin2020, Yu2021}. Another line of research has investigated the capabilities opened up by modern automatic differentiation (AD) and machine learning (ML) toolboxes \cite{Kandel2019,Seifert2021}. AD frameworks offer flexibility as they simply require the user to specify a desired forward model, typically specified in the form of a suitable loss function and possible regularizers. This approach is convenient for the user as it dispenses with the need to derive challenging gradient expressions, the latter of which oftentimes involve complex-valued (Wirtinger) derivatives \cite{Kreutz-Delgado2009, brandwood1983complex}. It is thus for instance straightforward to switch from one regularizer to another without analytically deriving and programming the underlying gradient expressions into the underlying software. ML approaches, in particular those based on neural networks, have been used to significantly speed up the reconstruction process, lower the sampling requirements in the raw data, and to embed denoising priors \cite{cherukara2020ai, aslan2021joint}. However, neural network approaches need to be trained based on data sets that have already been solved for the corresponding real space images. In reference \cite{cherukara2020ai} the neural network was trained based on the solution of an iterative solver. Moreover, training a neural network capable of solving ptychography problems is a memory-consumptive, large-scale computational challenge, that cannot be performed on small hardware architecture. We thus believe the need for memory-efficient but possibly slower iterative algorithms remains, despite the exciting possibilities opened up by neural networks \cite{harder2021deep}. From the above referenced work, we shortly describe some of the features of two prominent code projects, namely \emph{PtychoShelves} \cite{wakonig2020ptychoshelves} and \emph{PtyPy} \cite{Enders2016}, to illustrate some of the different design choices made in \emph{PtyLab}. 

\emph{PtychoShelves} \cite{wakonig2020ptychoshelves} is a Matlab-based software package for ptychography, designed with large-scale synchrotron facilities in mind. \textit{Shelves} refer to the modular coding framework representing bookshelves, from which desired \textit{books} (e.g., detector module, reconstruction engine etc.) can be taken out and inserted into the processing pipeline. To provide data handling across synchrotron facilities worldwide, \emph{PtychoShelves} supports commonly used X-ray detectors as well as instrument control software. Reconstructions are highly-optimized for Matlab-based GPU acceleration as well as CPU processing through reconstruction engines written in binary C++ code and parallelized through the OpenMP multiprocessing interface. The C++ code supports Difference Map \cite{Thibault2008} and Maximum Likelihood \cite{Thibault2012} engines, together with other features such as mixed state \cite{Thibault2013} and multi-slice ptychography \cite{Maiden2012multislice}. A wider range of reconstruction features are available through Matlab-based GPU engines, including an iterative least-squares solver for maximum-likelihood ptychography \cite{Odstrcil2018}, orthogonal-probe-relaxation \cite{Odstrcil2016OPR}, near-field ptychography \cite{Stockmar2013}, and position correction \cite{Maiden2012annealing}. 

\emph{PtyPy} \cite{Enders2016} is an open-source ptychography framework written in Python. It follows the Python coding style and is therefore modularized and object-oriented. The physical models are abstracted, which results in readable and concise code interfaces at a user level. The key element is the so-called POD class (probe, object, diffraction), which holds the access rule for a single position, object mode, and probe mode. For parallelization of the reconstructions, \emph{PtyPy} uses a message passing interface (MPI), which allows for flexible usage on standard office PCs, workstations, and clusters. MPI favors non-sequential reconstruction algorithms that can be parallelized (e.g. Difference Map \cite{Thibault2008} and Maximum Likelihood \cite{Thibault2012}). So far, a broad range of forward models are implemented (e.g. mixed state ptychography \cite{Thibault2013}, near-field ptychography \cite{Stockmar2013}, and lateral position correction \cite{Maiden2012annealing}). The \emph{PtyPy} framework is actively developed and novel features (e.g. GPU acceleration) are constantly added.

Both \emph{PtychoShelves} and \emph{PtyPy} are powerful ptychography data analysis platforms, but their design for high-performance computing poses an entry barrier for simple, one-off reconstructions in an academic-lab setting. In such cases, rapid code prototyping and ease-of-use can be more desirable than highly-optimized data handling and reconstructions.

Unlike CP, FP has not seen the same wide-spread use within research institutions that resulted in well-developed and maintained coding platforms. The existing open-source FP codes in \cite{Tian2014b,zheng2016fourier,bian2016motion,zuo2016adaptive,Aidukas2019a,zhou2020diffraction} exist mainly to supplement publications by providing only minimal working examples with limited functionality intended.  Recently an attempt has been made to provide a Matlab-based FP reconstruction platform \cite{rogalski2021fpm}, which among other features provides raw data denoising, GPU processing, and LED misalignment correction. 

Our goal here is to bridge the gap between CP and FP, thereby allowing for a cross-pollination of the two domains and a unified algorithmic framework. As compared to the above highlighted software packages \emph{PtyLab} is less focused on high performance and distributed computing, but puts emphasis on providing an ptychography ecosystem for researchers interested in rapid prototyping and exchanging algorithms - across modalities and programming languages.

\subsection{Outline}

In Section \ref{sec: equivalence of CP and FP} we revisit the idea of reciprocity, which formalizes the equivalence between CP and FP - the central idea for the unified software design in \emph{PtyLab}. Section \ref{sec: Code structure} details language-specific implementation details in Matlab, Python, and Julia. Section \ref{sec: Inverse Modeling} serves as a comprehensive overview of the available forward models and the corresponding optimization algorithms. Practical features for scan grid optimization are described in section \ref{sec: Scan grid optimization}. \emph{PtyLab} is released with various data sets and hands-on tutorials, which are described in section \ref{sec: Open experimental data and tutorials}. 

\section{Implications of reciprocity for ptychography}
\label{sec: equivalence of CP and FP}
\begin{figure}[htbp]
  \centering
  \includegraphics[width=\textwidth]{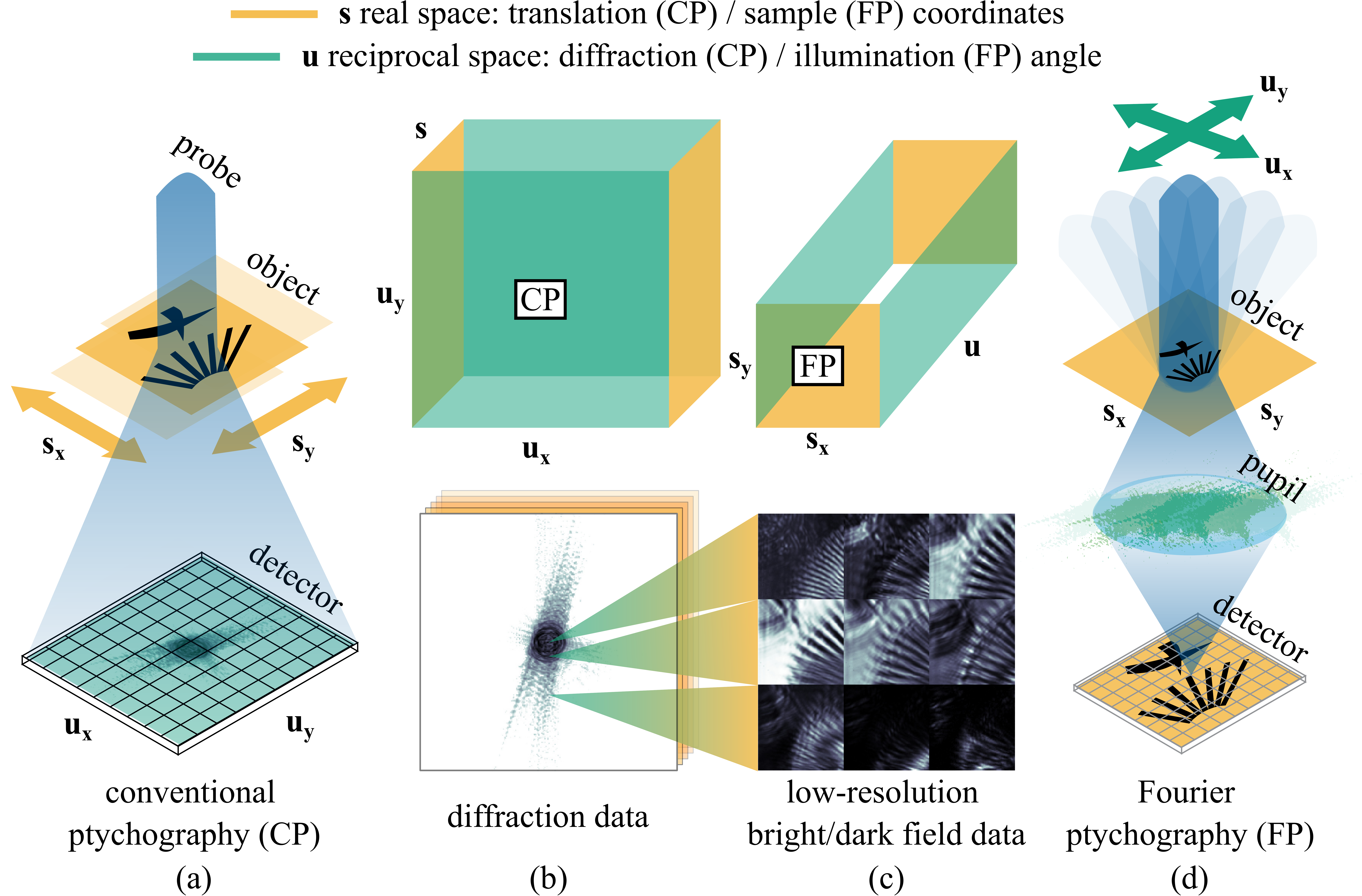}
\caption{ Illustration of the operation principle and equivalence of conventional and Fourier ptychography. (a) An object is laterally translated against a localized illumination profile. (b) In CP, the recorded data cube is a sequence of diffraction patterns, providing spatial frequency ($\boldsymbol{u}$) information for each scan position ($\boldsymbol{s}$). Each detector pixel alone contains a sequence real-space information that may be reshaped into a low-resolution real-space image of the sample. (c) In FP, the recorded data cube is a sequence of low-resolution bright and dark field image plane ($\boldsymbol{s}$) data corresponding to bandpass-filtered versions of the object spectrum. The shifts with respect to the pupil are controlled by shifting the illumination direction ($\boldsymbol{u}$). (d) Single-lens FP experimental configuration. Data in panel (b) and (c) from \cite{Loetgering2020helical} }
  \label{fig:reciprocity}
\end{figure}

One may think of the data sets recorded in ptychography in analogy to a musical score, where frequency information is prescribed at particular signatures in time. Once such a time-frequency, or phase-space, representation is given in form of a musical score, we can convert this information into either the time or the frequency domain. For example, we can digitally record a concert and Fourier transform the resulting signal. These processing steps would involve the temporal waveform and its frequency spectrum, respectively. Likewise, ptychography jointly samples real and reciprocal space representations of a signal, where for simplicity we ignore the additional complication of phase retrieval for the moment. The goal of ptychography is to convert partial phase-space information of a signal into a pure space or a pure spatial frequency representation. Physically, the phase-space description of ptychography \cite{Rodenburg1992} is intimately connected to the principle of reciprocity \cite{Born1999}, which states that by interchanging the illumination and detection direction in an optical system identical data sets can be observed. 

We would like to distinguish two types of reciprocity in ptychography. Type-I reciprocity refers to the ability to design separate CP and FP optical systems, both of which produce 4D data cubes which are essentially related by a phase space rotation \cite{Horstmeyer2014}. In addition, we define type-II reciprocity, which refers to the ability to algorithmically convert a 4D data cube from one domain to the other. Thus type-I reciprocity is essentially a statement about the ability to design different CP and FP hardware embodiments producing the same 4D experimental data cube. Type-II reciprocity is a matter of data processing: once a 4D data cube is measured in either a CP or a FP representation, it can be converted into the other respective domain and subsequently reconstructed. 

Figure \ref{fig:reciprocity} illustrates this idea of reciprocity in the context of ptychography. In CP (Fig.~\ref{fig:reciprocity}a), an object is laterally translated against a typically focused beam. This probe localizes the origin of the measured signal in real-space (scan coordinates $\boldsymbol{s}$). A sequence of diffraction patterns is captured on a pixelated detector, which is assumed here to be located in the far field (spatial frequency $\boldsymbol{u}$). Hence the data cube in CP consists of a sequence of angular scattering maps, each corresponding to a particular real-space specimen region localized to the spatial extent of the incident probe (Fig.~\ref{fig:reciprocity}b). In FP, an object is held at a fixed location under variation of the illumination angle (Fig.~\ref{fig:reciprocity}d). Thus the angular spectrum emanating from the sample is shifted over the finite-sized lens aperture. The pupil serves to bandpass filter the object's spatial frequency spectrum, resulting in a data cube consisting of dark and bright field images (Fig.~\ref{fig:reciprocity}c). Thus the data cube in FP consists of a sequence of real-space images, each corresponding to a particular reciprocal space portion of the object spectrum localized to the passband admitted by the pupil of the imaging system. In summary, both flavors of ptychography sample real and reciprocal space. 

We illustrate the aforementioned types of reciprocity in two ways: First, consider replacing the detector in the FP setup (Fig.~\ref{fig:reciprocity}d) with a pixelated light source. Turning on a single point source at a time and placing the detector in the far field of the sample, we can record a CP data set by scanning the point source location, as pointed out in a recent review article \cite{Rodenburg2019}. Of course, this hardware conversion ability faces practical limits imposed by the lens, which may cause a space-variant probe, but this complication is ignored here. Thus via hardware modification we can convert a CP experimental setup into and a FP system, which is a type-I reciprocity. Type-II reciprocity concerns the captured data itself and does not require any hardware modifications. It is possible to convert the measured data cube from one modality to the other. Suppose in Fig.~\ref{fig:reciprocity}a we scan the sample (for conceptual simplicity) on a regular raster grid and record a sequence of diffraction patterns. Then each pixel of the detector can be regarded as a traditional (single-pixel) scanning microscope data set. The data on each individual pixel may directly be reshaped into a two-dimensional real-space image, simply by permuting its dimension in correspondence with the scan trajectory. Practically, aperiodic translation trajectories and high NA effects require interpolation techniques. However, at low NA and using raster scan grids the data reshaping operation can be implemented with a single line of code (namely, a permute operation), converting for instance a CP data set into a sequence of low-resolution bright and dark field images, the latter of which constitutes the raw data for FP. While we described type-II reciprocity phenomenologically in this section, a mathematical proof of this conversion ability is provided in the appendix. The mathematical details also elucidate the correspondence between reconstructed quantities in CP and FP. We provide online tutorials that illustrate the conversion between CP and FP \cite{PtyLab}. 

The ability to convert CP and FP data has a bearing on the computational complexity of inversion algorithms underlying ptychography. Suppose we are given a CP data cube consisting of diffraction patterns with $U^2$ pixels at $S^2$ scan points. A single iteration of a parallelized ptychography solver (for example difference map \cite{Thibault2008}) requires us to numerically propagate exit waves from all scan positions to the detector plane and back. The Fourier transform operations involved will have a computational complexity of $\mathcal{O}\left[S^{2}\cdot U^{2}\cdot\log\left(U\right)\right]$ if we work in the CP domain, while it scales with $\mathcal{O}\left[U^{2}\cdot S^{2}\cdot\log\left(S\right)\right]$ if we convert the data into the FP domain. The difference in the log-terms can result in a practical speed up, provided that the number of detector pixels per dimension $U$ and the number of scan positions per dimension $S$ is largely different.

In summary, utilizing type-II reciprocity is the central motivation for the design of \emph{PtyLab}: CP and FP data can be converted into each other. A unified data analysis framework thus allows to migrate between the two modalities. A benefit of this data conversion ability is the applicability of diverse inversion algorithms and self-calibration routines in the domain in which they are most conveniently applied. Another benefit of reciprocity is the trade-off in computational complexity.

\section{Code Structure}
\label{sec: Code structure}
In this section we describe structural workflow in \emph{PtyLab}. Our overall goal is to provide a code that enables flow of algorithmic ideas and rapid prototyping beyond the boundaries of modality (CP/FP) and programming language (Matlab/Python/Julia). Thus collaborators with different programming language preferences or from different communities (e.g. synchrotron-based CP versus visible light FP) can easily exchange code, without being perfectly literate in the other programming language. This approach comes at the benefit of a unified structure and naming convention but at times at the expense of certain language-specific programming conventions. The following subsection describes the common structure independent of programming language. Subsequently, we address differences in the Matlab, Python, and Julia implementations. 

\subsection{Platform- and modality-independent workflow}

\begin{figure}[htbp]
  \centering
  \includegraphics[width=\textwidth]{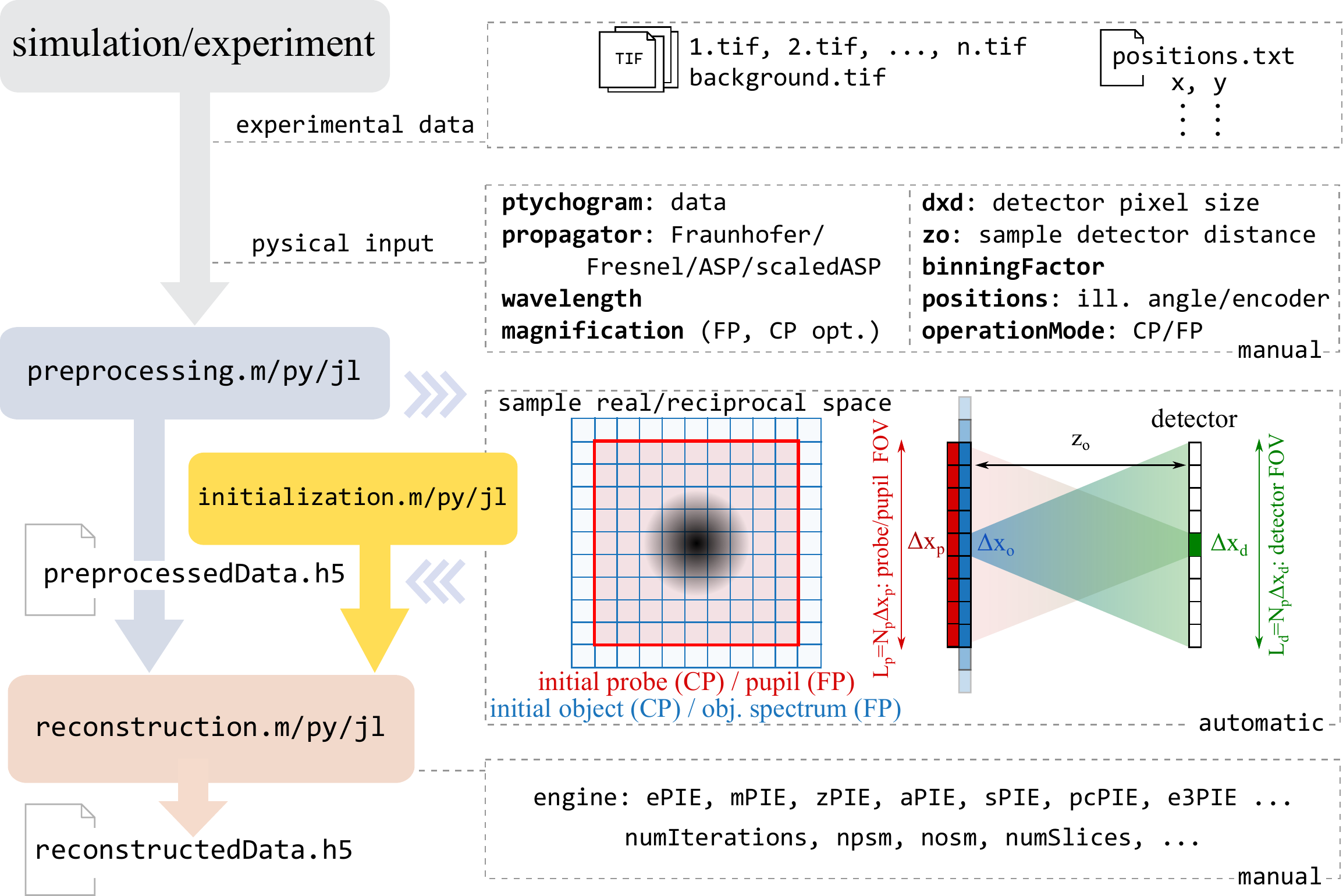}
\caption{ Workflow in \emph{PtyLab}. Experimental data is converted into a preprocessed hdf5 data set. The remaining parameters controlling algorithmic and monitoring behavior and required for reconstruction are set by an initialization routine. Various reconstruction engines and forward models can be chosen to analyze the preprocessed data. After the reconstruction is finished the reconstructed data is written into a final hdf5 file.}
  \label{fig:codeWorkflow}
\end{figure}

\begin{figure}[htbp]
  \centering
  \includegraphics[width=0.7\textwidth]{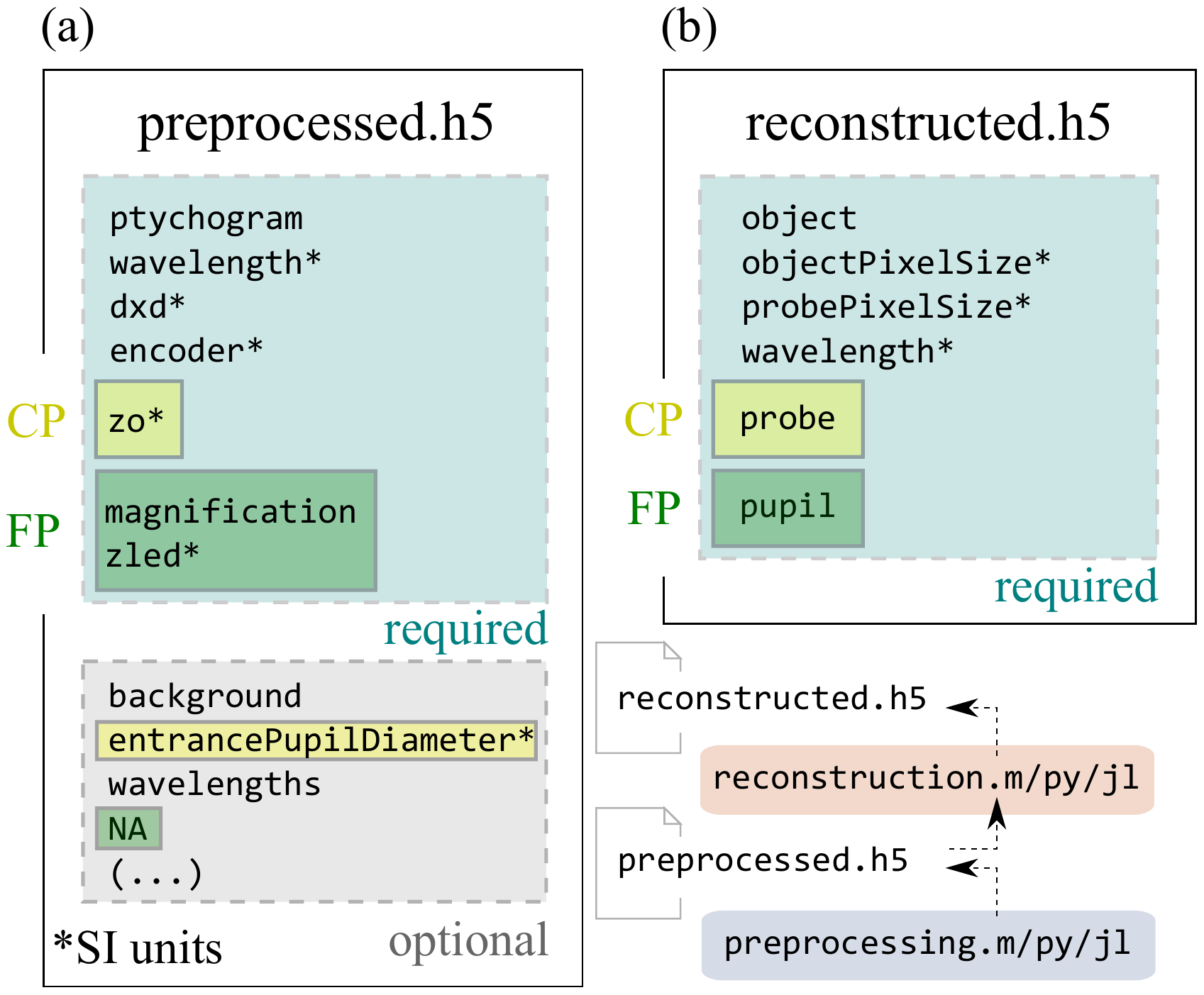}
\caption{ \emph{PtyLab} HDF file structure of preprocessed (a) and reconstructed (b) data. The orange boxes show the mandatory fields, with specific differences between CP (yellow) and FP (green) data. The grey box in panel (a) shows optional fields, that are nevertheless recommended. Both CP and FP reconstructions struggle to converge when background is not appropriately subtracted or accounted for in the forward model, subtraction being the easier route. Initial estimates for the probe diameter in CP and the pupil diameter in FP are recommended to be specified (both referred to as {\tt entrancePupilDiamater}). This can aid initial convergence in CP. Moreover, the circle fitting routine for position calibration in FP \cite{Eckert2018}, which is used in \emph{PtyLab}, requires an estimate of the pupil diameter. Arrays indicated with (*) are specified in SI units.}
  \label{fig:h5fileStructure}
\end{figure}

A high-level overview of the \emph{PtyLab}'s workflow is illustrated in Fig.~\ref{fig:codeWorkflow}. Assuming CP or FP data is stored in a local folder on the user's computer, the first step is preprocessing the data (see Fig.~\ref{fig:codeWorkflow}). Preprocessing converts the raw data into a \emph{PtyLab} class object. The user specifies physical input (illumination wavelength), hardware properties (detector pixel pitch, binning), and geometric parameters (sample-detector distance [CP], lens magnification [FP], sample scan trajectory [CP], illumination angles [FP]). In addition, the user specifies a forward model that describes the propagation from end to end (CP: probe to detector, FP: pupil to detector). The preprocessing pipeline then writes a single or multiple \emph{PtyLab} class objects into a hdf5 file (see Fig.~\ref{fig:h5fileStructure}) \cite{folk2011overview}. 

Second, the reconstruction script loads the preprocessed data. An initialization function generates uniform or radomized starting estimates for the probe~(CP) or pupil~(FP) and object~(CP) or object~spectrum~(FP). In each the probe/pupil, object/object~spectrum, and detector planes, meshgrids are calculated. These meshgrids depend on the specified physical, hardware, and geometrical parameters, as well as a forward model (propagator) that describes the mapping between probe~(CP) or pupil~(FP) and detector planes. A variety of propagation models can be specified, including angular spectrum (AS), scaled angular spectrum (SAS), Fresnel (Fresnel), Fraunhofer (Fraunhofer), and tilted Fresnel diffraction. The latter is relevant for non-coplanar reflection geometries and is typically performed only once on the raw diffraction data stack, provided no angle correction is performed (see subsection \ref{subsec: aPIE} for further details). Fraunhofer and Fresnel diffraction distinguish each other by an incorporation of a quadratic phase inside the propagation model. This quadratic phase may be absorbed into the probe/pupil function and can be compensated for post-reconstruction when a quantitative analysis of the reconstructed wavefront~(CP) or pupil~(FP) is of interest to the user.

\subsection{Matlab structure}

\begin{figure}[htbp]
  \centering
  \includegraphics[width=\textwidth]{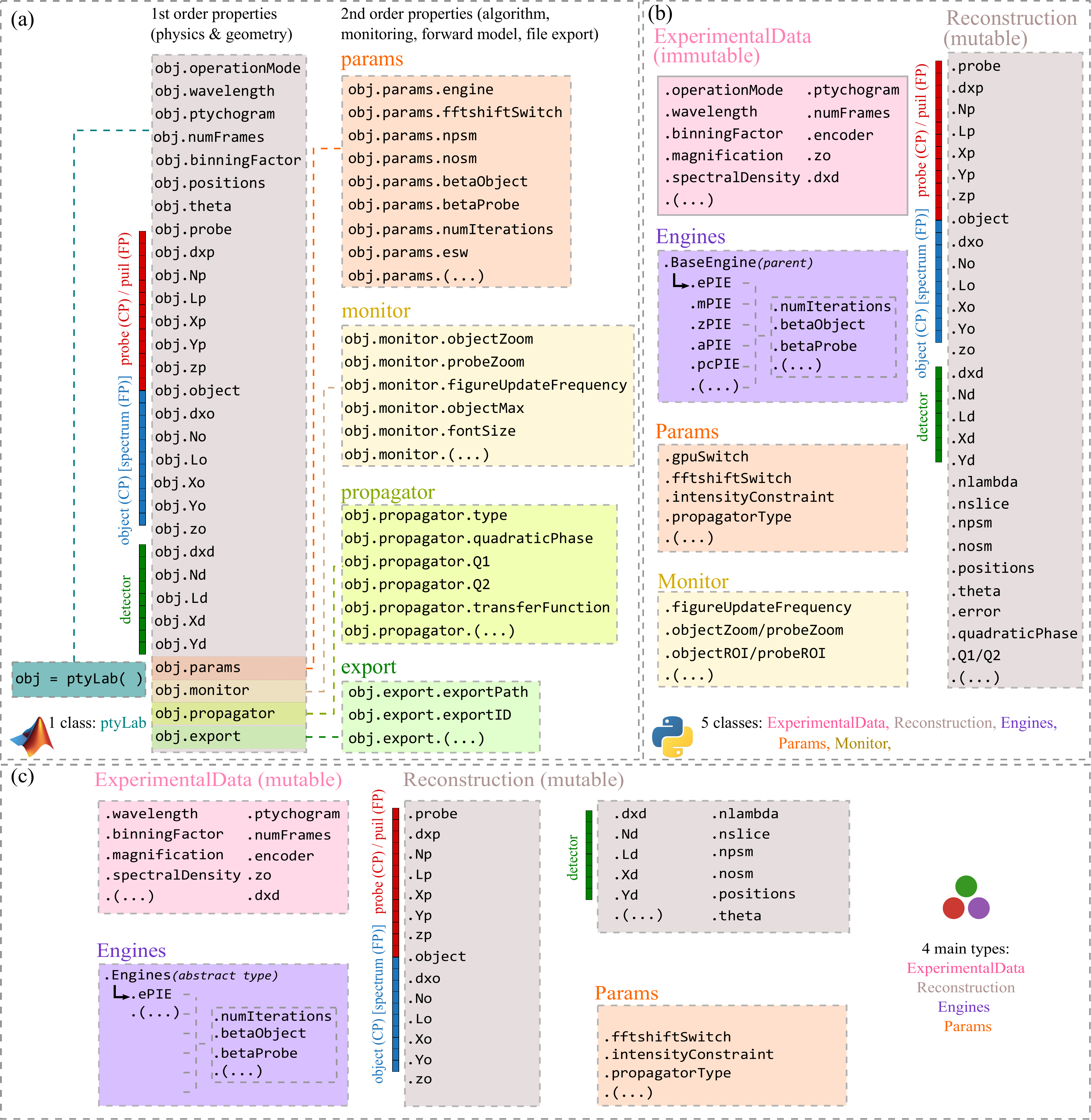}
\caption{The Matlab code structure (left) comprises a single class, which contains all field relevant for ptychographic data analysis.The Matlab class is organized into first- and second-order properties. First-order properties contain physical information (e.g. {\tt wavelength}) and geometrical parameters of the experiment. Second-order properties are mainly found in {\tt params}, which contains algorithmic properties (step sizes, number of iterations, etc.) that are optimized during data analysis. Other second-order properties comprise monitoring behaviour ({\tt monitor}), specification of the wave propagation model ({\tt propagator}), and input-output control ({\tt export}). The Python code (right) consists of five separate classes: {\tt ExperimentalData, Reconstruction, Params, Monitor, and Engines}. The Julia implementation consists of 4 main abstract types called {\tt ExperimentalData, Reconstruction, Params, and Engines}.}
  \label{fig:MatlabPythonClasses}
\end{figure}

The Matlab code structure is shown in Fig. \ref{fig:MatlabPythonClasses}a. Here an object of class \emph{PtyLab} is generated. Its first-order properties ({\tt obj."firstOrder"}) contain the physics as well as the geometry of the experiment. In addition, there are second-order properties ({\tt obj.params."secondOrder"}), which are algorithmic parameters ({\tt obj.params}), monitoring control ({\tt obj.monitor}), propagators as part of the forward model ({\tt obj.propagator}), and file export parameters ({\tt obj.export}). Certain notational conventions are noteworthy: the diffraction or image data is contained in {\tt obj.ptychogram}, a term borrowed from time-frequency analysis where the raw data is often referred to as \emph{spectrogram} \cite{brunton2022data}.  The dimensions of ({\tt obj.ptychogram}) are $({\tt y, x, numFrames})$, which is different from the Python convention (see subsection \ref{subsec:Python}). The order along the first two dimensions stems from Matlab's most convenient use when adhering to row-column convention. Similarly, {\tt obj.positions} follows row-column convention. 

\subsection{Python structure}

\label{subsec:Python}
The Python structure is similar in idea to the Matlab structure, but is designed with a stronger emphasis on modularity. 
As shown in Fig.~\ref{fig:MatlabPythonClasses}(b), the Python implementation contains five classes:  \texttt{ExperimentalData, Reconstruction, Monitor, Params}, and  \texttt{Engines}. Most but not all of these classes reflect second-order properties in the Matlab structure.
The  \texttt{ExperimentalData} class imports data from a preprocessed .hdf5 file, checks if all required parameters for a ptychographic reconstruction are included, and saves them into an instance that is immutable. 

The  \texttt{Reconstruction} class takes the  \texttt{ExperimentalData} instance as the input, and creates a mutable instance containing attributes that are optimized during a reconstruction process, e.g. the probe/pupil, and the object, as well as attributes that are related to the optimizable parameters, e.g. the error, the coordinates and meshgrids. 
Note that in the Python implementation, the probe/pupil and the object are set as 6D arrays with the fixed axes {\tt [nlambda, nosm, npsm, nslice, row, col]}, which are the number of wavelength, object state mixtures, probe state mixtures, slices (for multislice ptychography), and rows as well as columns. 

The  \texttt{Monitor} class is used to display a reconstruction process. Equivalent to its Matlab counterpart, one or two figures are created depending on the verbosity level set by users. A default figure shows the updated object, probe, and reconstruction error. An optional figure shows the comparison of the measured and estimated diffraction patterns. The update frequency of the plots can also be controlled by users ({\tt Monitor.figureUpdateFrequency}). 

The  \texttt{Params} class holds parameters that determine how a reconstruction is performed, for instance whether a reconstruction is carried out on a CPU or a GPU, the propagator type such as  \texttt{Fraunhofer, Fresnel}, (scaled) angular spectrum (\texttt{ASP, scaledASP}), etc., and whether the order of position iterations is random or sequential. Switches and parameters of various regularization types are also included in the \texttt{Params} instance, for example controlling how frequenctly orthogonalization is applied in the context of a mixed-states reconstruction.

The  \texttt{Engine} class consists of a  \texttt{BaseEngine} as a parent class, and other child engine classes, for instance ePIE \cite{Maiden2009}, mPIE \cite{Maiden2017}, zPIE \cite{Loetgering2020zPIE}, aPIE \cite{deBeurs2022apie}, and qNewton \cite{Yeh2015}. All four instances of \texttt{ExperimentalData, Reconstruction, Params}, and  \texttt{Monitor} are taken as inputs for a chosen engine, then get modified/updated by the engine, and can be passed to a different engine easily. Each engine stores its own attributes such as the number of iterations ({\tt numIteration}), and the update step sizes for the probe/pupil ({\tt betaProbe}) and object ({\tt betaObject}).

\subsection{Julia structure}

\textit{PtyLab.jl} is the most recent translation of \textit{PtyLab} to Julia. Due to the differences in Julia to Matlab and Python consequently small differences exist in the implementation but most of the common principles still hold. The amount of features is less than in the other two packages since its focus was on performance first.
The basis are four main types {\tt ExperimentalData, Reconstruction, Params, and Engines}. \texttt{Engines} is an abstract type which is subtyped (indicated by \texttt{<:}) by specific algorithms (such as \texttt{ePIE <: Engines}) as composite types. Via this mechanism, generic functions can be commonly used by all \texttt{Engines} solvers. However, Julia's multiple dispatch allows that different functionalities can be specified if they belong to a \texttt{ePIE} or a \texttt{zPIE} solver.
\texttt{Params} is a composite type storing second-order-properties.
Further, \texttt{ExperimentalDataCPM <: ExperimentalData} exists and similarly \texttt{ReconstructionCPM <: Reconstruction} to store the experimental data and the reconstruction data. During the iteration of the algorithms, the fields of \texttt{ReconstructionCPM} are allowed to change.
Julia's language features allow for a functional style of programming implying that memory buffers are not explicitly exposed to the user but instead are implicitly stored via closures.

\section{Inverse modeling}
\label{sec: Inverse Modeling}
\begin{figure}[htbp]
  \centering
  \includegraphics[width=\textwidth]{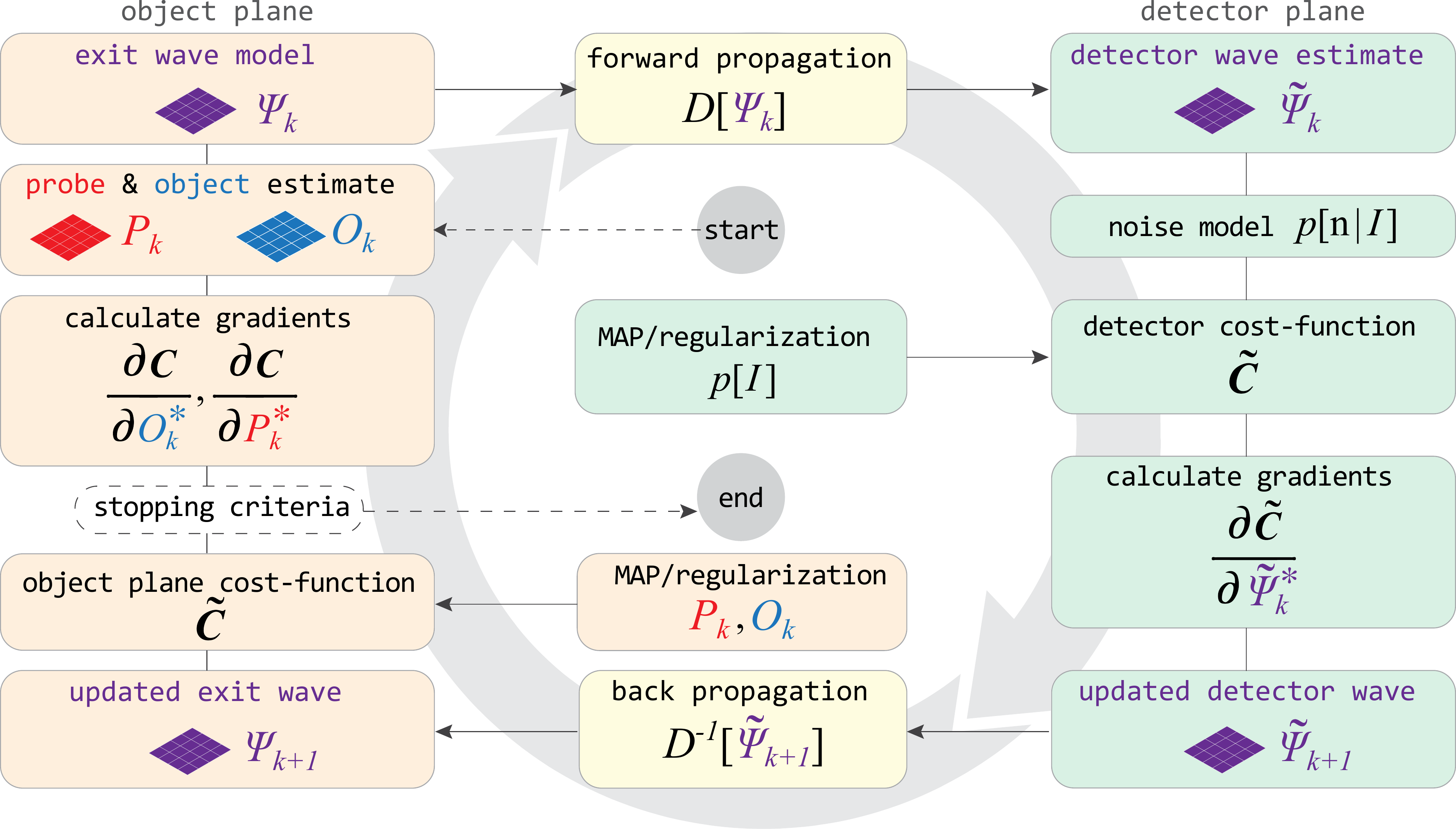}
\caption{Optimization workflow. The schematic illustrates the building blocks of a user-defined reconstruction routine in \emph{PtyLab}. In the object plane, the forward exit wave model as well as the inverse model for the object and probe gradients, subject to several regularization options, are specified. In the detector plane, the underlying noise model and various regularization options lead to an optimal update for the estimated detector wave. After initialization of the object and probe, the reconstruction engine iterates between the object and detector plane until the reconstruction error is low or other stopping criteria are satisfied. }
  \label{fig:optimizationWorkflow}
\end{figure}

The inverse modeling workflow in \emph{PtyLab} consists of several modular processing steps, which are shown in Fig.~\ref{fig:optimizationWorkflow}. All optimization algorithms in \emph{PtyLab} iterate between the object (CP) / object spectrum (FP)  plane (orange) and the detector plane (green), where the two planes are linked via a suitable propagation model (yellow). We subdivide this section into several parts, describing the individual modules that the user can stack up to build customized data analysis pipelines.

\subsection{Forward model}
In CP and FP the goal is to retrieve a wide-field high-resolution reconstruction of a sample of interest. In addition, the probe (CP) and pupil (FP) of the imaging system are recovered. A forward model links the estimated detector intensity $I$ to the object $O$ and probe $P$ (CP) or object spectrum $\tilde{O}$  and pupil $\tilde{P}$ (FP), 

\begin{equation}
\label{eqn:forwardModel}
    \begin{split}
        & I_{j}\left(\boldsymbol{q}\right)=\left|\mathcal{D}_{\boldsymbol{r}\rightarrow\boldsymbol{q}}\left[P\left(\boldsymbol{r}\right)\cdot O\left(\boldsymbol{r}-\boldsymbol{r}_{j}\right)\right]\right|^{2}  \;\; \textrm{(CP)} \\
        & I_{j}\left(\boldsymbol{r}\right)=\left|\mathcal{D}_{\boldsymbol{q}\rightarrow\boldsymbol{r}}\left[\tilde{P}\left(\boldsymbol{q}\right)\cdot\tilde{O}\left(\boldsymbol{q}-\boldsymbol{q}_{j}\right)\right]\right|^{2} \; \textrm{(FP)}.
    \end{split}
\end{equation}
Here $\mathcal{D}$ describes wave propagation between the sample~(CP)/pupil~(FP) and the detector plane, $\boldsymbol{r}$ refers to spatial coordinates, and $\boldsymbol{q}$ refers to reciprocal space coordinates. The index $j=1,...,J$ denotes scan position. For simplicity, we drop the coordinate dependence and use the CP notation throughout; the conversion to FP of all results discussed below is straightforward. The symbol $O$ refers to a particular object box of equal size as the probe FOV (compare red region in Fig.~\ref{fig:codeWorkflow}). The entire object field of view is denoted by $O_{\textrm{FOV}}$ (compare blue region in Fig.~\ref{fig:codeWorkflow}).

In the presence of noise in the observed signal, for instance caused by photoelectric conversion and read-out, we cannot expect to find a combination of sample and prope/pupil that exactly matches the recorded data. In what follows we therefore assume the measured data $m$ to arise as a probabilistic response to the true intensity $I$ incident on the detector and discuss several maximum likelihood estimation (MLE) models \cite{Dilanian2010, Godard2012, Thibault2012, Yeh2015, Odstrcil2018,Yan2020}. These MLE models aim at estimating the most likely combination of object and probe, a viewpoint extended by the addition of maximum a posteriori (MAP) estimation,  which originates from a Bayesian viewpoint and enables flexible embedding of regularization into the forward model \cite{Thibault2012, Pelz2017, Yan2020}. Before going into the details of inverse models, we briefly review the continuous and discrete viewpoints on optimization, which are both encountered in the ptychography optimization literature \cite{Guizar-Sicairos2008, Dilanian2010, Godard2012, Thibault2012,Thibault2013, Marchesini2013a, Yeh2015, Horstmeyer2015_ConvexRelaxation, Maiden2017, Odstrcil2018, Yan2020}.

\subsection{Inverse modeling}

\begin{figure}[htbp]
  \centering
  \includegraphics[width=0.8\textwidth]{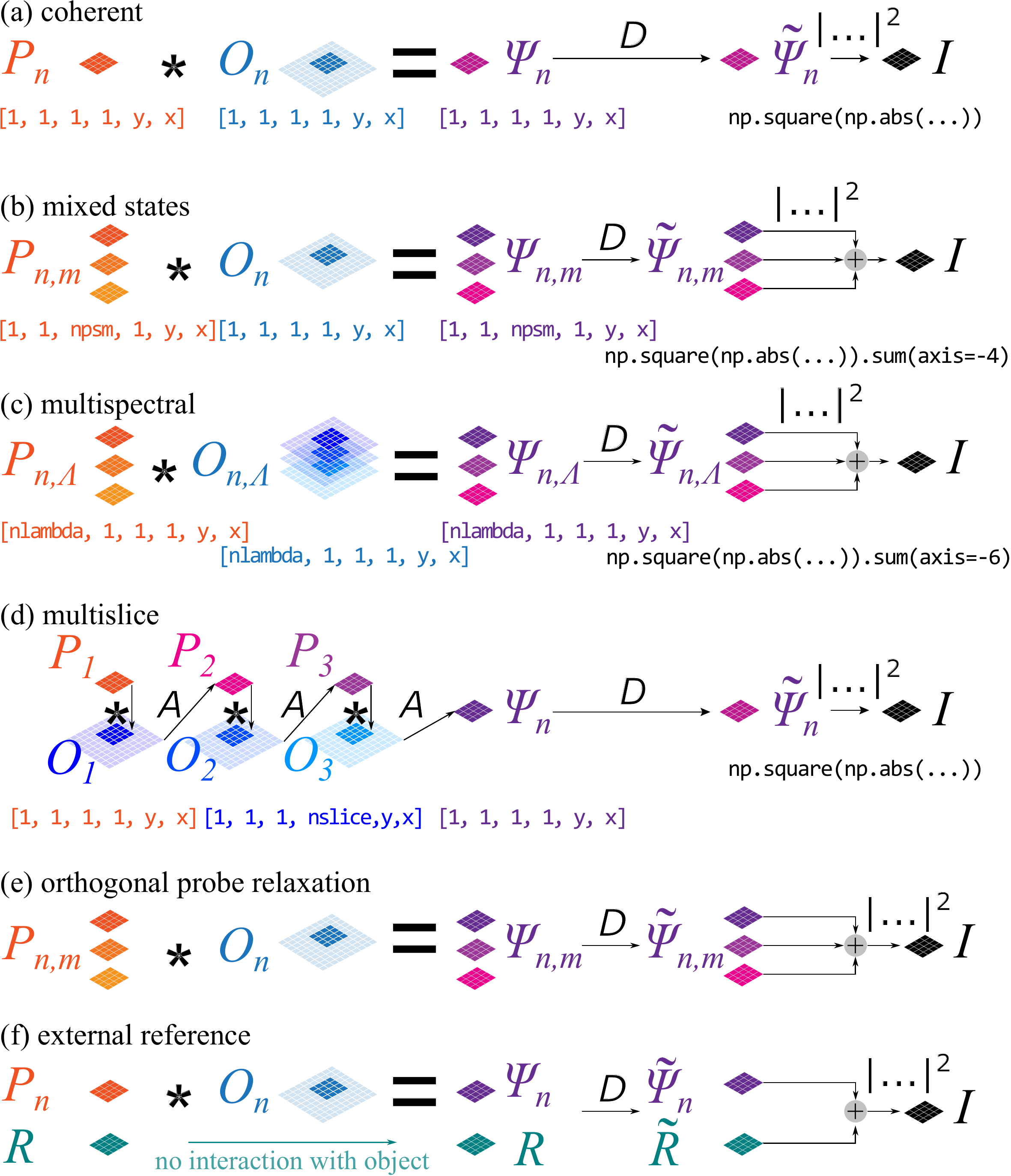}
\caption{ Selection of forward models implemented in \emph{PtyLab}: (a) The basic coherent diffraction model assumes the thin element approximation (TEA), where the exit wave $\psi_j$ is modeled as a product of probe $P$ and object box $O_j$ a scan position $j$. The exit wave is propagated into the detector plane via a suitable propagator $D$. (b) In mixed state ptychography the object interacts with $k$ mutually incoherent probe modes, giving rise to independent exit waves $\psi_{j,k}$. These exit waves are propagated into the detector plane and incoherently added to form the intensity forward model. (c) Multispectral ptychography. Here a polychromatic probe interacts with a dispersive object, both of which are functions of wavelength $\Lambda$. (d) In multislice ptychography the exit wave is modeled using the beam propagation method (BPM), which models a three-dimensional object to consist of several two-dimensional slices (index $s$). Inside each slice the TEA is used, while the propagation between slices is carried out via angular spectrum propagation $A$. (e) Orthogonal probe relaxation can model scan position dependent probes $P_j$ as a linear combination of mutually coherent orthogonal basis modes $U_k$. (f) A coherent external reference wave can be added to the forward model.}
  \label{fig:forwardModels}
\end{figure}

In this section we review various forward models implemented in \emph{PtyLab}. A summary of these forward models is given in Fig.~\ref{fig:forwardModels}. We first describe general techniques to tackle the inverse problem underlying ptychography. Subsequently, we detail the individual solvers that allow the user to build and invert modular forward models.

\subsubsection{The continuous viewpoint}

In the continuous viewpoint, we aim to minimize a cost functional 
\begin{equation}
    C=\int\mathcal{C}\left(\boldsymbol{r},f\left(\boldsymbol{r}\right),\boldsymbol{g}\right)d\boldsymbol{r},
\end{equation}
where the functional density $\mathcal{C}$ is a real-valued, non-negative and at least once differentiable function. We use the abbreviation $\boldsymbol{g}=\nabla f\left(\boldsymbol{r}\right)$ for notational brevity in the equations to follow. For real-valued functions $f$ minimizing the cost functional $C$ is equivalent to solving the Euler-Lagrange equation \cite{Gbur2011}

\begin{equation}
    \frac{\partial\mathcal{C}}{\partial f}-\textrm{div}_{\boldsymbol{g}}\left(\frac{\partial\mathcal{C}}{\partial\boldsymbol{g}}\right)=0,
\end{equation}
where $\textrm{div}_{\boldsymbol{g}}$ is the divergence with respect to the third (vector-valued) input of $\mathcal{C}$. For complex-valued $f$, we may solve two separate Euler-Lagrange equations for the two degrees of freedom of $f$, for instance its real and imaginary parts. We can save some work if we regard $f$ and its complex conjugate $f^{*}$ as the degrees of freedom to solve for \cite{brandwood1983complex,Frieden1991}. In the particular case that the cost density $\mathcal{C}$ is symmetric, i.e.

\begin{equation}
    \frac{\partial\mathcal{C}\left(\boldsymbol{r},f,\boldsymbol{g}\right)}{\partial f^{*}}=\left(\frac{\partial\mathcal{C}\left(\boldsymbol{r},f,\boldsymbol{g}\right)}{\partial f}\right)^{*},
\end{equation}
it suffices to solve a single Euler-Lagrange equation  \cite{Kreutz-Delgado2009}
\begin{equation}
\label{eqn:ELE}
    \frac{\partial\mathcal{C}}{\partial f^{*}}-\textrm{div}_{\boldsymbol{g}}\left(\frac{\partial\mathcal{C}}{\partial\boldsymbol{g}^{*}}\right)=0.
\end{equation}
If Eq.~\ref{eqn:ELE} is not amenable to a direct solution, we can iteratively solve it by seeking the steady state solution of the diffusion equation \cite{Rudin1992, Benyamin2020}
\begin{equation}
\label{eq: time-ELE}
    \frac{\partial f}{\partial t}=-\alpha\left[\frac{\partial\mathcal{C}}{\partial f^{*}}-\textrm{div}_{\boldsymbol{g}}\left(\frac{\partial\mathcal{C}}{\partial\boldsymbol{g}^{*}}\right)\right],
\end{equation}
where $\alpha$ controls the diffusion step size. Approximating the time derivative by finite differences, we may rewrite Eq. \ref{eq: time-ELE} as
\begin{equation}
\label{eq:functionalGradientDescent}
    f_{k+1}=f_{k}-\alpha\left[\frac{\partial\mathcal{C}}{\partial f_{k}^{*}}-\textrm{div}_{\boldsymbol{g}_{k}}\left(\frac{\partial\mathcal{C}}{\partial\boldsymbol{g}_{k}^{*}}\right)\right],
\end{equation}
where $k$ denotes iteration. We refer to this update as \emph{functional gradient descent}. Under some circumstances to be discussed below the divergence term vanishes. In this case we identify this update with the Wirtinger derivative previously discussed in \cite{Thibault2012, Candes2015, Xu2018}. However, we will make use of regularizers which require this more general update rule.

\subsubsection{The discrete viewpoint}
The discrete viewpoint is used when considering inverse problems over sampled functions. In this case, we oftentimes wish to minimize the sum of squares cost function

\begin{equation}
\label{eqn:discreteProblem}
    \mathcal{C}=\sum_{k}\lambda_{k}\left\Vert \boldsymbol{A}_{k}f-\tilde{\psi_{k}}\right\Vert _{2}^{2},
\end{equation}
where $\left\Vert \ldots\right\Vert _{2}$ denotes the L2 norm. Here $\boldsymbol{A}_{k}$ is a matrix and $f$ is a vector, which are compatible in dimensions. The gradient to this problem is given by \cite{Boyd2019}
\begin{equation}
\label{eqn:discreteGradient}
    \frac{\partial\mathcal{C}}{\partial f^{*}}=\sum_{k}\lambda_{k}\boldsymbol{A}_{k}^{\dagger}\left(\boldsymbol{A}_{k}f-\tilde{\psi}_{k}\right),
\end{equation}
where the matrix $\boldsymbol{A}_{k}^{\dagger}$ is the conjugate transpose of $\boldsymbol{A}_{k}$. We may iteratively solve the original problem in Eq.~\ref{eqn:discreteProblem} using gradient descent
\begin{equation}
\label{eqn:discreteGradientUpdate}
    f_{n+1}=f_{n}-\alpha\sum_{k}\lambda_{k}\boldsymbol{A}_{k}^{\dagger}\left(\boldsymbol{A}_{k}f-\tilde{\psi}_{k}\right).
\end{equation}
A non-iterative solution is formally obtained by setting the gradient in Eq.~\ref{eqn:discreteGradient} to zero and solving for $f,$
\begin{equation}
\label{eq:LSQ}
    f=\left(\sum_{k}\lambda_{k}\boldsymbol{A}_{k}^{\dagger}\boldsymbol{A}_{k}\right)^{-1}\left(\sum_{k}\lambda_{k}\tilde{\psi}_{k}\right),
\end{equation}
which is referred to as the \emph{least squares solution}. We note that the transition between the continuous and discrete viewpoints is seamless, provided that the signals of interest are bandlimited. In this case one may switch between the continuous and discrete viewpoints by adequate sampling and interpolation \cite{Bracewell2003}. 

\subsection{Maximum likelihood (MLE) estimation}
We now discuss models for the detector noise commonly used in ptychography. Two particularly prominent models that have been addressed \cite{Dilanian2010, Godard2012, Thibault2012, Yeh2015, Odstrcil2018} are the Poisson likelihood 

\begin{equation}
\label{eqn:Poisson}
    p\left[m\left|I\right.\right]=\frac{\left(I+b\right)^{m+b}}{\left(m+b\right)!}\exp\left[-\left(I+b\right)\right] \; \; \; \textrm{(shifted Poisson)} 
\end{equation}
and the Anscombe likelihood  
\begin{equation}
\label{eqn:Anscombe}
    p\left[m\left|I\right.\right]=\exp\left[-\left(\sqrt{m+b}-\sqrt{I+b}\right)^{2}\right] \; \; \; \textrm{(Anscombe)}, 
\end{equation}
where $I=\tilde{\psi}^{*}\tilde{\psi}$ is the estimated intensity, , $\tilde{\psi}=\mathcal{D}\left(P\cdot O\right)$ (CP, cf. Eq.~\ref{eqn:forwardModel}; similarly for FP), and $m$ is the measured intensity. In both cases, the offset term $b$ is typically not made explicit in the literature, although needed to prevent division by zero in the maximum likelihood gradients (cf. Eqs. \ref{eqn:PoissonGradient} and \ref{eqn:AnscombeGradient}). In case of the Poisson likelihood the additional term $b$ has previously been used to account for detection models that contain mixed Poissonian and Gaussian noise contributions and is also referred to as the shifted Poisson approximation \cite{Chakrabarti2012, Chouzenoux2015, Ikoma2018}. The Anscombe model transforms Poisson-distributed data, which exhibits exposure dependent shot noise, into variance-stabilized data with uniform uncertainty across variable exposure, which is the basis for robust denoising \cite{Makitalo2011}. However, while the Anscombe transform has been noted to stabilize variance, it can introduce bias and tends to underestimate the true mean of the signal in the limit of low exposure \cite{murtagh1995image}. We recommend setting the offset to at least $b=1$ to prevent division by zero in the gradient descent update rules derived below. 

Computing the Wirtinger derivatives of the negative log-likelihood $\mathcal{L}=-\sum_{\boldsymbol{q}}\log\left(p\left[m\left|I\right.\right]\right)$ of Eqs. \ref{eqn:Poisson} and \ref{eqn:Anscombe} results in \cite{Godard2012,Thibault2012,Yeh2015, Odstrcil2018}
\begin{equation}
\label{eqn:PoissonGradient}
    \frac{\partial\mathcal{L}}{\partial\tilde{\psi}^{*}}=\left(1-\frac{m+b}{I+b}\right)\tilde{\psi} \; \; \; \textrm{(shifted Poisson gradient)} 
\end{equation}
and 
\begin{equation}
\label{eqn:AnscombeGradient}
    \frac{\partial\mathcal{L}}{\partial\tilde{\psi}^{*}}=\left(1-\sqrt{\frac{m+b}{I+b}}\right)\tilde{\psi} \; \; \; \textrm{(Anscombe gradient)}. 
\end{equation}

It appears to us that the Anscombe forward model is used by the vast majority in the ptychography literature. Although the Poisson distribution works in practice, we have observed that the Anscombe model is more robust in practical data analysis. One has to keep in mind that the Poisson model assumes that the photoelectric counting distribution's mean equals its variance and is only valid for shot-noise limited data - a somewhat restrictive assumption considering the manifold fluctuations that are present in typical experiments, including partial spatial and temporal coherence effects as well as detector read out. We note that other models for the statistics of photoelectic counting distributions have been proposed in the literature, albeit to our knowledge they not yet have been used for ptychography. Noteworthy is the negative binomial distribution
\begin{equation}
\label{eq: negative-binomial}
    p\left[m\left|I\right.\right]=\frac{\left(m+M-1\right)!}{m!\left(M-1\right)!}\cdot\frac{I^{m}\cdot M^{M}}{\left(I+M\right)^{m+M}}, \; \; \; \textrm{(negative-binomial)} 
\end{equation}
which was first derived by Mandel \cite{mandel1959fluctuations}. The parameter $M$ counts the degrees of freedom in the detected light. For an integration time $T$ much longer than the coherence time $\tau_c$ the degrees of freedom can be estimated as $M=T/\tau_c$ \cite{Goodman2015}. Notice that this number does not have to be an integer and one can simply replace the factorials in Eq.~\ref{eq: negative-binomial} by gamma functions. However, leaving the factorials it is easy to see that for large $M$ the negative binomials approximately equals a Poisson distribution. In the other extreme case that $M=1$, the negative binomial distribution degenerates into a geometric distribution, which is the noise distribution for thermal light measured at time scale approaching the coherence time \cite{mandel1959fluctuations}. Thus by varying $M$ one can parameterize between the degree to which the Poisson model is relaxed. The gradient of the negative log-likelihood of the negative binomial distribution is given by
\begin{equation}
    \frac{\partial\mathcal{L}}{\partial\tilde{\psi}^{*}}=\left[\frac{m+M}{I+M}-\frac{m}{I}\right]\tilde{\psi}, \; \; \; \textrm{(negative-binomial gradient)} 
\end{equation}
which has the desired property that it vanishes for $I=n$, similar to Eqs.~\ref{eqn:PoissonGradient} and \ref{eqn:AnscombeGradient}. For large $M$ the first fraction approaches one and we recover the Poisson gradient (compare Eq.~\ref{eqn:PoissonGradient}). It is an interesting possibility left for future studies to test the performance of a generalized Anscombe gradient of the form
\begin{equation}
    \frac{\partial\mathcal{L}}{\partial\tilde{\psi}^{*}}=\left[\sqrt{\frac{m+M+b}{I+M+b}}-\sqrt{\frac{m+b}{I+b}}\right]\tilde{\psi}, \; \; \; \textrm{(generalized Anscombe gradient)} 
\end{equation}
which results from taking the square root of the fractions in the negative-binomial gradient. In the limit of large $M$ the latter gradient approaches the Anscombe gradient.
 
\subsection{Maximum a posteriori (MAP) estimation}

The MLE approach in the previous subsection can be extended by MAP estimation, which introduces prior knowledge into the reconstruction process. In MAP the detector intensity is regarded as a random variable with underlying probability density $p\left[I\right]$. Since the detector intensity $I=\tilde{\psi}^{*}\tilde{\psi}$ is a function of the real space object and probe, namely $\psi\left(\boldsymbol{r}\right)=P\left(\boldsymbol{r}\right)\cdot O\left(\boldsymbol{r}-\boldsymbol{r_{j}}\right)$ and $\tilde{\psi}\left(\boldsymbol{q}\right)=\mathcal{F}_{\boldsymbol{r}\rightarrow\boldsymbol{q}}\left[\psi\left(\boldsymbol{r}\right)\right]$, MAP opens up a convenient way to formulate and impose constraints in the inverse problem underlying ptychography. The optimization problem is then 

\begin{equation}
    \textrm{argmax}_{P,O}\prod_{\boldsymbol{q}}p\left[m\left(\boldsymbol{q}\right)\left|I\left(\boldsymbol{q}\right)\right.\right]p\left[I\left(\boldsymbol{q}\right)\right],
\end{equation}
which is equivalent to
\begin{equation}
\label{eqn:MAP}
    \textrm{argmin}_{P,O}\sum_{\boldsymbol{q}}-\log\left(p\left[m\left(\boldsymbol{q}\right)\left|I\left(\boldsymbol{q}\right)\right.\right]\right)-\log\left(p\left[I\left(\boldsymbol{q}\right)\right]\right).
\end{equation}

\subsubsection{Proximal detector updates}

The detector update can be restricted to small step sizes  by introducing the proximal prior
\begin{equation}
    p\left[I\right]=\exp\left[-\lambda\left(\sqrt{\left|\tilde{\psi}_{n+1}\right|^{2}}-\sqrt{\left|\tilde{\psi}_{n}\right|^{2}}\right)^{2}\right], \; \; \; \textrm{(proximal prior)}  
\end{equation}
where $\lambda$ controls how strong changes in the magnitude of the estimated detector wave are penalized between successive iterations $n$ and $n+1$. Inserting this into Eq.~\ref{eqn:MAP} together with the shifted Poisson (Eq.~\ref{eqn:Poisson}) and Anscombe (Eq.~\ref{eqn:Anscombe}) likelihood, we get the cost functions

\begin{equation}
\label{eqn:costProximalPoisson}
    \mathcal{C}=\left|\tilde{\psi}_{n+1}\right|^{2}-m\cdot\log\left(\left|\tilde{\psi}_{n+1}\right|^{2}\right)+\lambda\left(\sqrt{\left|\tilde{\psi}_{n+1}\right|^{2}}-\sqrt{\left|\tilde{\psi}_{n}\right|^{2}}\right)^{2} \; \; \; \textrm{(proximal Poisson)}  
\end{equation}
and
\begin{equation}
\label{eqn:costProximalAnscombe}
    \mathcal{C}=\left(\sqrt{\left|\tilde{\psi}_{n+1}\right|^{2}}-\sqrt{m}\right)^{2}+\lambda\left(\sqrt{\left|\tilde{\psi}_{n+1}\right|^{2}}-\sqrt{\left|\tilde{\psi}_{n}\right|^{2}}\right)^{2}. \; \; \; \textrm{(proximal Anscombe)}  
\end{equation}
These updates are referred to as \emph{proximal}. A large value of the tuning parameter $\lambda$ forces the updated wave $\tilde{\psi}_{n+1}$ to remain in the proximity of the previous estimate $\tilde{\psi}_{n}$. Intuitively, the updates along the gradient directions in Eqs.~\ref{eqn:PoissonGradient} and \ref{eqn:AnscombeGradient} enforce the magnitude of the updated wave to be equal to the measured data, either in intensity or modulus for the Poisson and Anscombe gradients, respectively. However, due to noise, sequential update schemes can only be as certain as the noise in a single diffraction pattern permits. Proximal gradients incorporate the memory of previous updates and do not naively accept the update suggested by the data. The gradient direction suggested by the data is followed in case that the deviation from the current estimate is small. It is conjectured here that this incorporates dose fractionation effects into ptychography, resulting in superior signal-to-noise in the reconstruction. This is supported by previous reports that observed improved reconstruction quality using proximal gradients for Gerchberg-Saxton type phase retrieval \cite{Soulez2016} and ptychography \cite{Marchesini2013a,Yan2020}. 
The cost functions above (Eqs.~\ref{eqn:costProximalPoisson} and \ref{eqn:costProximalAnscombe}) result in the update steps 
\begin{equation}
    \tilde{\psi}_{n+1}=\frac{m+\lambda\left|\tilde{\psi}_{n}\right|^{2}}{\left(1+\lambda\right)\left|\tilde{\psi}_{n}\right|^{2}}\tilde{\psi}_{n} \; \; \; \textrm{(proximal Poisson update)},
\end{equation}
and
\begin{equation}
    \tilde{\psi}_{n+1}=\frac{\sqrt{m}+\lambda\left|\tilde{\psi}_{n}\right|}{\left(1+\lambda\right)\left|\tilde{\psi}_{n}\right|}\tilde{\psi}_{n}. \; \; \; \textrm{(proximal Anscombe update)}.  
\end{equation}
We note that the proximal Poisson update is different from the result in \cite{Soulez2016}, since here the prior is Gaussian in modulus, while in the related work the prior is Gaussian in intensity. The approach reported here avoids the need to solve a cubic polynomial to compute the proximal Poisson update. 

\subsubsection{Proximal probe and object updates via (e)PIE}
While in the previous section a proximal update step has been discussed in the detector update step, we may impose a similar type of regularization for the probe and object update, which has been shown to result in the ptychographic iterative engine (ePIE) \cite{Thibault2013}. This derivation is reviewed here from the discrete viewpoint outlined above. Considering the cost function
\begin{equation}
    \label{eq:ePIEcost}
    \mathcal{C}	=\left\Vert \mathcal{A}O_{n+1}-\tilde{\psi}\right\Vert _{2}^{2}+\alpha\left\Vert \Gamma O_{n+1}-\Gamma O_{n}\right\Vert _{2}^{2},
\end{equation}
the first term is the overlap constraint of ptychography while the second penalizes the step size in the search direction for the object $O$, which here is a vector. The operator matrix $\mathcal{A}=\mathcal{D}\mathcal{P}$ contains both the propagator $\mathcal{D}$ and a diagonal probe matrix $\mathcal{P}$, which act on the object. The matrix $\Gamma$ allows for regularization of the object. The detector wave $\tilde{\psi}$ is the obtained from the detector update step, as described in the previous subsections, where we have omitted an index to focus on the update of the object. Noting that $\mathcal{A}^{\dagger}\mathcal{A}=\mathcal{P}^{\dagger}\mathcal{D}^{\dagger}\mathcal{D}\mathcal{P}=\mathcal{P}^{\dagger}\mathcal{P}$, application of the least square solution in Eq.~\ref{eq:LSQ} results in 
\begin{align}
\label{eqn:kernelForm}
    O_{n+1}&=\left(\mathcal{A}^{\dagger}\mathcal{A}+\alpha\Gamma^{\dagger}\Gamma\right)^{-1}\left(\mathcal{A}^{\dagger}\tilde{\psi}+\alpha\Gamma^{\dagger}\Gamma O_{n}\right)
	\nonumber \\&=\left(\mathcal{A}^{\dagger}\mathcal{A}+\alpha\Gamma^{\dagger}\Gamma\right)^{-1}\left(\mathcal{A}^{\dagger}\tilde{\psi}-\mathcal{A}^{\dagger}\mathcal{A}O_{n}+\mathcal{A}^{\dagger}\mathcal{A}O_{n}+\alpha\Gamma^{\dagger}\Gamma O_{n}\right)
	\nonumber \\&=\left(\mathcal{A}^{\dagger}\mathcal{A}+\alpha\Gamma^{\dagger}\Gamma\right)^{-1}\mathcal{A}^{\dagger}\left(\tilde{\psi}-\mathcal{A}O_{n}\right)+O_{n}
	\nonumber \\&=\left(\mathcal{P}^{\dagger}\mathcal{P}+\alpha\Gamma^{\dagger}\Gamma\right)^{-1}\mathcal{P}^{\dagger}\left(\mathcal{D}^{\dagger}\tilde{\psi}-\mathcal{P}O_{n}\right)+O_{n}.
\end{align}
The particular choice 
\begin{equation}
    \Gamma_{\textrm{PIE}}=\textrm{diag}\left[\left(\frac{\max\left(\left|P\right|\right)}{\alpha\beta\left|P\right|}\left(\left|P\right|^{2}+\epsilon\right)-\frac{\left|P\right|^{2}}{\alpha}\right)^{1/2}\right]
\end{equation}
results in the original version of the ptychographic iterative engine (PIE) \cite{Faulkner2004a,Rodenburg2004}, namely
\begin{equation}
    O_{n+1}=O_{n}+\beta\frac{\left|P_{n}\right|}{\max\left(\left|P_{n}\right|\right)}\frac{P_{n}^{*}}{\left|P_{n}\right|^{2}+\epsilon}\left(\psi-P_{n}\cdot O_{n}\right).
\end{equation}
Later the extended ptychographic iterative engine (ePIE) was proposed \cite{Maiden2009}, which uses the regularization matrix \cite{Thibault2013}
\begin{equation}
    \Gamma_{\textrm{ePIE}}=\textrm{diag}\left[\left(\frac{\max\left(\left|P\right|^{2}\right)}{\alpha\beta}-\frac{\left|P\right|^{2}}{\alpha}\right)^{1/2}\right],
\end{equation}
resulting in the ePIE update \cite{Maiden2009}
\begin{equation}
    O_{n+1}=O_{n}+\beta\frac{P_{n}^{*}}{\max\left|P_{n}\right|^{2}}\left(\psi-P_{n}O_{n}\right).
\end{equation}
An intuition of the difference between PIE and ePIE can be obtained in the limit of $\epsilon\rightarrow0$ and $\beta=1$, for which we get
\begin{equation}
    \Gamma_{\textrm{PIE}}=\textrm{diag}\left[\left(\frac{\left|P\right|^{2}}{\alpha}\left(\frac{\max\left(\left|P\right|\right)}{\left|P\right|}-1\right)\right)^{1/2}\right]
\end{equation}
and 
\begin{equation}
    \Gamma_{\textrm{ePIE}}=\textrm{diag}\left[\left(\frac{\max\left(\left|P\right|^{2}\right)}{\alpha}-\frac{\left|P\right|^{2}}{\alpha}\right)^{1/2}\right].
\end{equation}
$\Gamma_{\textrm{PIE}}$ is small, when $\left|P\right|$ approaches $\max\left(\left|P\right|\right)$ or $0$. Thus PIE allows object updates for locations where the probe amplitude is small or large. In contrast, $\Gamma_{\textrm{ePIE}}$ is small, only when $\left|P\right|$ approaches $\max\left(\left|P\right|\right)$. Thus ePIE allows object updates only for locations where the probe amplitude is large. The derivation of the probe update is similar, resulting in a joint optimization of $P$ and $O$. The robustness of ePIE is attributed to the penalized step size at low probe intensities. However, in FP the large dynamic range of the object spectrum can cause problems in conjunction with ePIE. The pupil would be updated only in the center of k-space, where the object spectrum exhibits values close to its maximum amplitude. High illumination angles produce dark field images, which have a reduced signal-to-noise ratio as compared to bright field images from lower illumination angles. In CP the illuminating beam is always aligned with the detector resulting in images with similar signal-to-noise ratio as compared between scan positions. For this reason, we use the PIE-type update by default in \emph{PtyLab} for FP and the ePIE-type update for CP data analysis.  Other choices of regularization can be embedded into the reconstruction routine by changing the weight matrix $\Gamma$. We note in passing that the PIE-type update rule has been identified as a quasi-Newton algorithm \cite{Yeh2015}.

\subsubsection{Tikhonov regularization}

A popular idea in solving inverse problems is Tikhonov regularization. The general idea is to add an additional term to the cost function, which penalizes variations in the object,
\begin{equation}
\label{eq:TikhonovCost}
    \mathcal{C}=\left|P\cdot O-\psi\right|^{2}+\lambda\left|\nabla_{x,y} O\right|^{2}.
\end{equation}
We emphasize that the regularization term in Eq.~\ref{eq:ePIEcost} in the previous subsection penalizes fluctuations at for a fixed object pixel between successive iterations. In this subsection the regularization term in Eq.~\ref{eq:TikhonovCost} penalizes fluctuations between neighboring object pixels. Applying functional gradient descent, as described by Eq.~\ref{eq:functionalGradientDescent}, to the cost in Eq.~\ref{eq:TikhonovCost} gives
\begin{equation}
\label{eq:TikhonovUpdate}
   O_{n+1}=O_{n}-\alpha P^{*}\left(\psi-P\cdot O_{n}\right)+\alpha\lambda\Delta O_{n},
\end{equation}
where 
\begin{equation}
    \Delta O_{n}=\frac{\partial^{2}O_{n}}{\partial x^{2}}+\frac{\partial^{2}O_{n}}{\partial y^{2}}=-\mathcal{F}^{-1}\left(\left[\left(2\pi q_{x}\right)^{2}+\left(2\pi q_{y}\right)^{2}\right]\mathcal{F}\left(O_{n}\right)\right)
\end{equation}
is the Laplacian. Subtracting the Laplacian in Eq.~\ref{eq:TikhonovUpdate} removes high-frequency components from the object. Thus introducing Tikhonov regularization results in a low-pass filter smoothing the object \cite{Thibault2012, Stockmar2015}. While the smoothing operation is effective in preventing noise in the object reconstruction, it results in unwanted loss of high resolution features in the reconstruction. An alternative regularization with more favorable edge preservation properties is discussed in the next subsection. 

\subsubsection{Total variation regularization}
Total variation (TV) regularization penalizes changes in the image while to a certain degree preserving edge features \cite{Rudin1992}. The corresponding cost function can be approximated by 
\begin{equation}
    \mathcal{C}=\left|P\cdot O-\psi\right|^{2}+\lambda\sqrt{\left|\nabla O\right|^{2}+\epsilon}.
\end{equation}
Applying functional gradient descent (Eq.~\ref{eq:functionalGradientDescent}) gives
\begin{equation}
\label{eq:TV}
    O_{n+1}=O_{n}+\alpha P^{*}\left(\psi-P\cdot O_{n}\right)+\lambda\,\textrm{div}\left(\frac{\nabla O_{n}}{\sqrt{\left|\nabla O_{n}\right|^{2}+\epsilon}}\right),
\end{equation}
where $\textrm{div}$ denotes divergence. Equation~\ref{eq:TV} is applied to the complex-valued object. Alternative implementations \cite{Tanksalvala2021} have reported application of a TV prior to the real and imaginary part of the object separately, which is not equivalent to our implementation due to the nonlinearity of the TV regularizer.

\subsection{Momentum acceleration (mPIE and mqNewton)}
The momentum-accelerated ptychographic iterative engine (mPIE) \cite{Maiden2017} is the standard solver used for CP in \emph{PtyLab}. In mPIE a predefined number of ePIE iterations $T$ is carried out, after which the search direction is complemented by a momentum term $\nu$ updating the entire object field of view $O_{\textrm{oFOV}}$,
\begin{align}
    \nu_{n}&=\eta\cdot\nu_{n-T}+O_{n,\textrm{oFOV}}-O_{n+1-T,\textrm{oFOV}} \\
    O_{n+1,\textrm{oFOV}}&=O_{n,\textrm{oFOV}}+\eta\cdot\nu_{n}.
\end{align}
Here $\eta$ is a damping term that is set to 0.7 by default\cite{Maiden2017}. Similar to conjugate gradient solvers \cite{Thibault2012, qian2014efficient, Odstrcil2018}, the momentum term accelerates the search direction and prevents zigzag motion towards the optimum. We emphasize that $O_{n+1,\textrm{oFOV}}$ in this subsection denotes the entire probe field of view, while in other subsection $O$ is an object box of the same size as the probe window.

While addition of momentum is typically done for another regularized version of the PIE-type family of algorithms (rPIE, see \cite{Maiden2017}), it can complement any of the existing reconstruction engines including PIE. To avoid naming ambiguities, addition of momentum to PIE will be referred to as momentum accelerated quasi-Newton ({\tt mqNewton}), which we often use as an FP solver.

\section{Robust inverse models}
In the foregoing section, we have reviewed the basic inverse models underlying ptychography. However, oftentimes a variety of systematic errors are present in the experimental data that requires  more robust inverse models. This is the case when the data is corrupted by for example partial spatial as well as partial temporal coherence, when the illumination wavefront profile is unstable throughout the scan, and when the scan positions are imprecisely known. In what follows, we discuss robust forward models that account for and mitigate the aforementioned sources of error.

\subsection{Mixed States}

In mixed state ptychography \cite{Thibault2013} the intensity in the detector plane is modeled as 
\begin{equation}
    \label{eq:mixed state forward model}
    I=\sum_{k}\tilde{\psi}_{k}^{*}\tilde{\psi}_{k},
\end{equation}
where the index $k$ discerns mutually incoherent signal contributions (also known as \emph{mixed states}). Inserting this into Eqs.~\ref{eqn:Poisson} and \ref{eqn:Anscombe} and calculating the Wirtinger derivative with respect to each $\tilde{\psi}_{k}$, we get the gradients
\begin{equation}
    \label{eq:mixed state detector update}
    \frac{\partial\mathcal{L}}{\partial\tilde{\psi}_{k}^{*}}=\left[1-\left(\frac{n+b}{I+b}\right)^{p}\right]\tilde{\psi}_{k},
\end{equation}
where $p=1$ for the Poisson model and $p=1/2$ for the Anscombe model.

The real space cost function for mixed state ptychography is 
\begin{align}
    \mathcal{C}=\sum_{k}\left|P_{n+1,k}\cdot O_{n}-\psi_{k}\right|^{2}+\lambda_{P}\sum_{k}\left|P_{n+1,k}-P_{n,k}\right|^{2}+\lambda_{O}\left|O_{n+1}-O_{n}\right|^{2},
\end{align}
where the particular choices $\lambda_{P}=\frac{1}{\beta}\max\left|O_{n}\right|^{2}-\left|O_{n}\right|^{2}$ and $\lambda_{O}=\frac{1}{\beta}\max\left(\sum_{k}\left|P_{n,k}\right|^{2}\right)-\sum_{k}\left|P_{n,k}\right|^{2}$ and setting the Wirtinger derivatives with respect to $P_{n+1}$ and $O_{n+1}$ to zero lead to 
\begin{align}
    P_{n+1,k}&=P_{n,k}+\frac{\beta}{\max\left|O_{n}\right|^{2}}O_{n}^{*}\left(\psi_{k}-P_{n,k}\cdot O_{n}\right) \\
    O_{n+1}&=O_{n}+\frac{\beta}{\max\left(\sum_{k}\left|P_{n,k}\right|^{2}\right)}\sum_{k}P_{n,k}^{*}\left(\psi_{k}-P_{n,k}\cdot O_{n}\right),
\end{align}
which is a modified version of ePIE for the case of mixed states, as first derived in \cite{Thibault2013}. In \emph{PtyLab} a snapshot singular value decomposition \cite{brunton2022data} is used to orthogonalize the probe states during the reconstruction process, which allows for their interpretation as orthogonal modes of the mutual intensity of a partially coherent field provided that other decoherence effects are absent in the experimental data \cite{Loetgering2018, Rohrich2021}. It is noteworthy that multiple object states can be reconstructed as well \cite{Thibault2013}, provided the illumination is fully coherent, which otherwise leads to ambiguities \cite{Li2016}.

\subsection{Multispectral ptychography}
In multispectral ptychography \cite{Batey2014, dong2014spectral, Zhang2016, song2020super, Loetgering2021, goldberger2021spatiospectral, yao2021broadband} a light source of multiple individual spectral lines or a continuous spectrum is used. Because different colors are mutually incoherent, the detector update is identical to mixed state ptychography (compare Eq.~\ref{eq:mixed state detector update}), but the index $k$ now denotes wavelength instead of spatial mode. The differences between mixed state and multispectral ptychography lie in the real space updates and in the propagator between the sample and the detector plane. In \emph{PtyLab} we minimize the following cost function

\begin{align}
    \mathcal{C}=&\sum_{k=1}^{K}\left|P_{n+1,k}\cdot O_{n,k}-\psi_{k}\right|^{2}+\sum_{k=1}^{K}\lambda_{P,k}\left|P_{n+1,k}-P_{n,k}\right|^{2} \\&+\sum_{k=1}^{K}\lambda_{O,k}\left|O_{n+1,k}-O_{n,k}\right|^{2}+\mu\sum_{k=2}^{K-1}\left|2O_{n+1,k}-O_{n,k+1}-O_{n,k-1}\right|^{2}
\end{align}
where $\mu$ is a user defined parameter that enforces similarity between adjacent spectral reconstructions and
\begin{align}
    \lambda_{P,k}=\frac{1}{\beta}\max\left|O_{n,k}\right|^{2}-\left|O_{n,k}\right|^{2} \\
    \lambda_{O,k}=\frac{1}{\beta}\max\left|P_{n,k}\right|^{2}-\left|P_{n,k}\right|^{2}.
\end{align}
These cost functions result in the updates
\begin{align}
    P_{n+1,k}&=P_{n,k}+\frac{\beta}{\max\left|O_{n,k}\right|^{2}}O_{n}^{*}\left(\psi_{k}-P_{n,k}\cdot O_{n}\right) \\
    O_{n+1,k}&=\frac{\gamma}{\gamma+2\mu\beta}O_{n,k}+\frac{\beta P_{n,k}^{*}\left(\psi_{k}-P_{n,k}\cdot O_{n,k}\right)+\beta\mu\left(O_{n,k+1}+O_{n,k-1}\right)}{\gamma+2\mu\beta},
\end{align}
where $\gamma=\max\left|P_{n,k}\right|^{2}$. At the boundaries of the spectral range ($k=1$ and $k=K$) the object updates are carried out without influence from adjacent spectral channels,
\begin{equation}
    O_{n+1,k}=O_{n,k}+\frac{\beta}{\max\left|P_{n,k}\right|^{2}}P_{n,k}^{*}\left(\psi_{k}-P_{n,k}\cdot O_{n,k}\right).
\end{equation}
The latter update also results for the special case $\mu=0$. In this case the spectral channels are only coupled by the incoherent model for the detector intensity (Eq.~\ref{eq:mixed state forward model}). In the presence of spectral regularization ($\mu\neq0$) we do not need a priori knowledge about the spectral weights of the incident beam \cite{Loetgering2021}. In the original work proposing multispectral ptychography no spectral regularization of adjacent channels was used. Instead the spectrum of the incident polychromatic probe was known a priori and used as a constraint in the optimization routine \cite{Batey2014}. 

The second difference between mixed state and multispectral ptychography is the propagation model. Other code projects, for instance \emph{PtyPy} \cite{Enders2016}, use zero padding to model the wavelength dependence of far-field wave propagation. In \emph{PtyLab} the pixel size of a monochromatic wave can be scaled by using two-step propagators ({\tt scaledASP}), which omits the need for spectrally dependent zero padding of the exit wave. For details the reader is referred to the supplementary information of \cite{Loetgering2021}. 

\subsection{Multislice ptychography (e3PIE)}

In multislice CP \cite{Maiden2012multislice} the specimen is modeled by a stack of 2D slices. The beam propagation method (BPM) \cite{cowley1957scattering} is used as a forward model for the exit wave. In each of the slices the thin element approximation is assumed to be valid. The cascade of multiplication with each slice and subsequent propagation enables the BPM to model multiple forward scattering effects. A basic version of multislice CP (termed e3PIE) is implemented in \emph{PtyLab}. For details, the reader is referred to the original work by Maiden et al. \cite{Maiden2012multislice} and subsequent work \cite{Godden2014, Tsai2016a}. We have not yet implemented multislice FP \cite{Tian2015} in the current version of \emph{PtyLab}, although such an enigine may come in future releases. 

\subsection{Orthogonal probe relaxation (OPR)}

A basic version of orthogonal probe relaxation (OPR) \cite{Odstrcil2016OPR} is implemented in \emph{PtyLab}. Instead of sharing the same probe across all scan positions in CP, as done for example in simple engines such as ePIE, OPR relaxes the requirement for a stable probe. The exit waves from different scan positions are used to estimate a low-rank basis, which seeks to model probe variations that occurred during a full scan, for example caused by pointing instability of the source. For details, the reader is referred to the original work by Odstrcil et al. \cite{Odstrcil2016OPR}. We note that OPR can be combined with mixed states, as recently described by Eschen et al. \cite{eschen2022material}. 

\begin{figure}[htbp]
    \centering
    \includegraphics[width=0.8\linewidth,trim={3cm 0 3cm 0}]{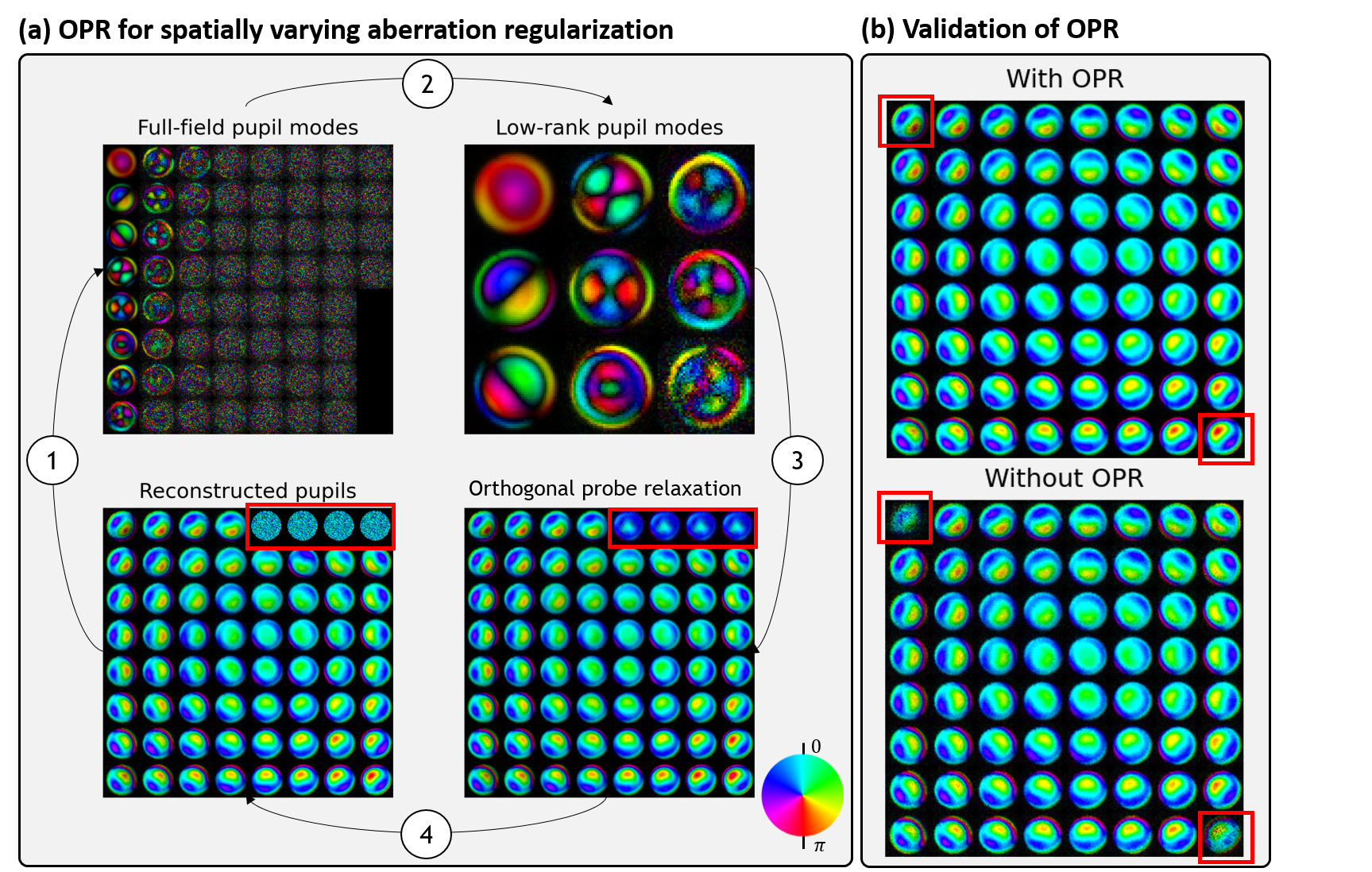}
    \caption{(a) Orthogonal probe relaxation scheme in FP. In step 1, the reconstructed pupils at a given iteration are factorized using SVD to produce an orthogonal full-field aberration basis. In step 2, the low-rank representation is obtained by eliminating modes with low-contributions to actual aberrations. This way, noise and errors are eliminated. In step 3, the low-rank full-field aberration reconstruction is performed from the low-rank basis. In this example, it can be seen that noisy pupils (top right corner) were replaced with a better pupil estimate. The whole process ensures that pupil aberrations remain well conditioned. (b) Experimental validation of pupil initialisation\cite{Aidukas2021}, where the use of OPR resulted in stable pupil reconstruction at the corners (indicated by red boxes).}
    \label{fig:OPR_FP}
\end{figure}

With regard to FP, to our knowledge OPR has not been applied, although this may be an interesting approach to effectively model space-variant pupil functions. The latter is typically achieved by partitioning a larger field of view into a set of smaller sub-regions, each of which may be subject to different pupil aberrations. This approach, known as embedded pupil recovery (EPRY) \cite{Ou2014}, has to date essentially remained the only model for space-variant pupil aberrations in FP. However, because EPRY requires the reconstruction of a separate pupil for each sub-region, the model requires many degrees of freedom and ignores that adjacent sub-regions are unlikely to have strongly differing pupil aberrations. OPR could be a promising candidate to robustify EPRY in future FP applications for spatially varying aberration reconstruction. An example of the use of OPR in FP is shown in  Fig.\ref{fig:OPR_FP}.  The image FOV was split into small segments, each  with its own unique pupil function and OPR was used to impose a low-rank consistency constraint on all the pupil functions. In applying OPR to FP adjacent pupils share information and poor convergence of some isolated field segments can be avoided (compare highlighted red boxes in Fig.~\ref{fig:OPR_FP}). For further implementation details the reader is referred to \cite{Aidukas2021}.

\subsection{Subsampling (sPIE)}

\begin{figure}[htbp]
  \centering
  \includegraphics[width=0.8\textwidth]{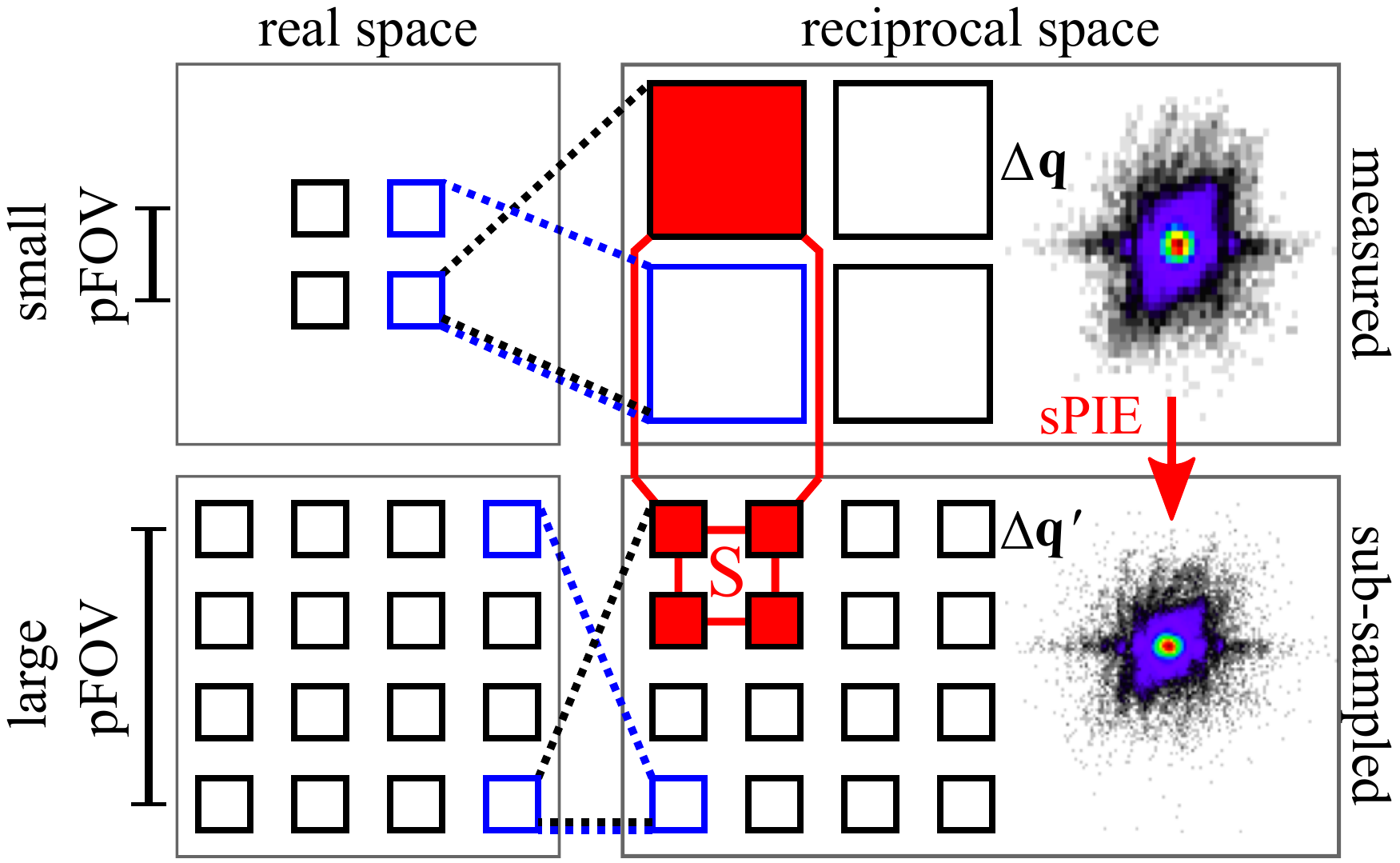}
\caption{ Illustration of sPIE in CP. Assuming a far-field diffraction geometry, the real and reciprocal space sampling conditions are inversely proportional (indicated by the dashed lines): A small probe field of view (pFOV) in CP requires only coarse detector pixels $\Delta\boldsymbol{q}$. Conversely, if the physical probe wavefront extends over a larger region, the observed data is undersampled (top row). In such situations, the detector pixels need to be sub-divided into smaller pixels $\Delta\boldsymbol{q}'$ (bottom row). This allows to extend the numerical pFOV to be larger than the physical probe size. The resulting constraint in the forward model is that the sum over the intensities over a set $\boldsymbol{S}$ of sub-sampled pixels equals the corresponding observation over the same region in the observed data with a coarse sampling grid.}
\label{fig: sPIE}
\end{figure}

In some applications it is challenging to sufficiently sample the captured detector signal. For example, in EUV ptychography generating a highly focused probe is oftentimes restricted by the available hardware \cite{Loetgering2022Advances}. In other applications, such as near-field ptychography, the detector pixel size can be a limiting factor \cite{xu2020super}. In both situations one may attempt to solve for the probe and object in CP (or the pupil and object spectrum in FP) from undersampled measurements, being too coarse to oversample the diffraction data. In principle, the detrimental effect of undersampling can be compensated by high overlap. In the context of CP this technique is known as reciprocal space upsampling and was first demonstrated by Batey et al. \cite{Batey2014a} (where the algorithm was named \emph{sPIE}). In the latter work, the captured ptychography measurements were deliberately undersampled by means of binning, but the original oversampled data could be recovered thanks to the high scan grid overlap in the captured data. We later generalized this principle to arbitrary sensing matrices that are not necessarily a result of an operation equivalent to binning, but that could result from any sensing architecture that compresses multiple, not necessarily neighboring pixels into a smaller data cube \cite{Loetgering2017a}. In such situations one seeks to minimize the cost function 
\begin{equation}
    \mathcal{L}=\left\Vert \sqrt{\boldsymbol{S}\left|\tilde{\psi}\right|^{2}}-\sqrt{I}\right\Vert _{2}^{2},
\end{equation}
where $\boldsymbol{S}$ is a sensing matrix representing, for example, downsampling or any other detection scheme. Gradient descent on this cost function results in a modified intensity projection given by
\begin{equation}
\label{eq:sPIE}
    \tilde{\psi}_{n+1}=\tilde{\psi}_{n}\boldsymbol{S}^{T}\sqrt{\frac{I}{\boldsymbol{S}\left|\tilde{\psi}_{n}\right|^{2}}},
\end{equation}
where $\boldsymbol{S^T}$ is the transpose of the sensing matrix. For the special case of $\boldsymbol{S}$ being a downsampling operation $\boldsymbol{S^T}$ is an upsampling operation (compare Fig.~\ref{fig: sPIE}). In this case, Eq.~\ref{eq:sPIE} modifies the estimated detector wave by multiplying it with the upsampled version of the ratio between the measured intensity $I$ (already downsampled) and the downsampled estimated intensity $\boldsymbol{S}\left|\tilde{\psi}_{n}\right|^{2}$. This principle was also used by Xu et al. who reported sub-sampled near-field ptychography \cite{Xu2018}.

\subsection{Lateral position correction (pcPIE)}

In ptychography the scan positions may not be accurately known, for example in the case of a low-precision scanning stage \cite{Guizar-Sicairos2008, Maiden2012annealing, Beckers2013, Hessing2016}. The wrong estimation of the scan positions will cause errors and artifacts during the stitching of the object patches into the large object field of view. In \emph{PtyLab} a momentum-accelerated version of a cross-correlation-based lateral position correction algorithm is used \cite{Zhang2013}. The rationale of this position correction is based on the observation that, at iteration $n+1$ of the reconstruction procedure, the object patch estimate at scan position $j$ is slightly shifted towards its true position. This shift is detected and used to update the scan grid by maximizing the cross correlation
\begin{equation}
    C_{n,j}(\Delta\boldsymbol{r})=\sum_{\boldsymbol{r}}O_{n,j}^{*}(\boldsymbol{r}-\Delta\boldsymbol{r})\cdot O_{n+1,j}(\boldsymbol{r}-\Delta\boldsymbol{r})
\end{equation}
with respect to the shift $\Delta\boldsymbol{r}$. In practice, we shift the object patch of the iteration $n$ one pixel in all directions (horizontally, vertically and diagonally) and compute the centre of mass of the cross correlation 
\begin{equation}
    \Delta_{n,j}=\frac{\sum_{\Delta\boldsymbol{r}}\Big(|C_{n,j}(\Delta\boldsymbol{r})|-\left\langle |C_{n,j}(\Delta\boldsymbol{r})|\right\rangle \Big)\Delta\boldsymbol{r}}{\sum_{\boldsymbol{r}}\left|O_{n,j}(\boldsymbol{r})\right|^{2}},
\end{equation}
where the brackets $\left\langle \ldots\right\rangle$ denote an average over all shift pixels, and then estimate the position gradient $d_{n,j}$ using
\begin{equation}
    d_{n,j}=\alpha\cdot\Delta_{n,j}+\beta\cdot d_{n-1,j}
\end{equation}
The updated scan position at iteration $n+1$ is
\begin{equation}
    \boldsymbol{r}_{n+1,j}=\boldsymbol{r}_{n,j}-d_{n}.
\end{equation}
Default values are $\alpha = 250$, $\beta = 0.9$, and $d_0=0$.

\subsection{Reflection ptychography with angle calibration (aPIE)}
\label{subsec: aPIE}

In reflection ptychography the sample plane and the detector are non-coplanar. Assuming far-field diffraction for simplicity, the captured data is related to the specimen exit wave by a Fourier transformation plus an additional coordinate transformation \cite{Patorski1983, Seaberg2014}. An inverse coordinate transformation can be applied to the captured raw data to simplify the forward model. However, this operation requires accurate knowledge of the angle between the optical axis and the specimen surface normal. If this angle is not calibrated within a fraction of a degree, the reconstruction quality can suffer notably. We have recently presented an algorithm for angular auto-calibration in reflection ptychography (aPIE), which is part of \emph{PtyLab}. aPIE uses a heuristic strategy to  estimate the unknown angle within an iteratively shrinking search interval. For details the reader is referred to \cite{deBeurs2022apie}.

\subsection{Axial position correction (zPIE)}
\label{subsec: zPIE}

Similarly to pcPIE (position correction) and aPIE (angle correction in reflection ptychography) another self-calibration algorithm provided as part of \emph{PtyLab} is zPIE, which can be used to estimate the sample-detector distance \cite{Loetgering2020zPIE}. The main idea is that when the sample-detector distance is miscalibrated, the reconstructed object oftentimes exhibits characteristics of a slightly defocused inline hologram, including ringing at edges. An autofocus metric based on total variation (TV) is then used to calibrate the correct sample detector distance. We observed that TV-based autofocusing performs best in the near-field on binary specimens, although it can also be used on biological specimens. Other choices of autofocusing metrics can easily be implemented by the user, if the TV-based sharpness metric fails \cite{fonseca2016comparative}. 

\subsection{Ptychography combined with an external reference beam}

We recently reported ptychographic optical coherence tomography, which combines full field frequency-domain OCT with ptychography \cite{Du2021}. In the latter work there was no need for an external reference wave, as common in OCT applications. Instead the reference was provided from a direct surface reflection of the sample itself. Thus the technique can principally be applied to the short-wavelength regime, where providing an external reference comes with extra experimental challenges. However, in the visible and infrared spectral range a reference wave is readily provided and can make POCT more convenient. Providing an external reference wave in ptychography requires adjustments to the forward model. In this case, we seek to minimize the cost function density
\begin{equation}
    \mathcal{L}=\left[\sqrt{\left|\tilde{\psi}+\tilde{\rho}\right|^{2}}-\sqrt{I}\right]^{2},
\end{equation}
where $\tilde{\rho}$ denotes a coherent external reference wave. All other quantities are the same as defined as in previous sections. Using gradient descent with a unit step size in conjunction with Wirtinger derivatives, we obtain updates for both the wave diffracted from the specimen
\begin{equation}
    \tilde{\psi}_{n+1}=\left(\tilde{\psi}_{n}+\tilde{\rho}_{n}\right)\sqrt{\frac{I}{\left|\tilde{\psi}_{n}+\tilde{\rho}_{n}\right|^{2}}}-\tilde{\rho}_{n}
\end{equation}
and the external reference wave
\begin{equation}
    \tilde{\rho}_{n+1}=\left(\tilde{\psi}_{n}+\tilde{\rho}_{n}\right)\sqrt{\frac{I}{\left|\tilde{\psi}_{n}+\tilde{\rho}_{n}\right|^{2}}}-\tilde{\psi}_{n}.
\end{equation}

We note that the mathematical structure of external reference beam ptychography opens up a trivial ambiguity. Suppose that the triplet of probe $P$, object $O$, and reference $\tilde{\rho}$ yields the observed intenisty $I$, i.e.
\begin{equation}
    I=\left|\tilde{\psi}+\tilde{\rho}\right|^{2}	=\left|\mathcal{F}\left(P\cdot O\right)+\tilde{\rho}\right|^{2}
	=\left|\mathcal{F}\left(P\right)\otimes\mathcal{F}\left(O\right)+\tilde{\rho}\right|^{2}.
\end{equation}
Then it immediately follows that the triplet of probe $P$, object $-O$, and reference $\tilde{P}+\tilde{\rho}$ is also a solution, since
\begin{align}
    \left|\tilde{\psi}+\tilde{\rho}\right|^{2}	&=\left|\mathcal{F}\left(P\cdot\left[1-O\right]\right)+\tilde{\rho}\right|^{2} \\
	&=\left|\tilde{P}+\mathcal{F}\left(P\right)\otimes\mathcal{F}\left(-O\right)+\tilde{\rho}\right|^{2} \\
	&=\left|\mathcal{F}\left(P\right)\otimes\mathcal{F}\left(-O\right)+\tilde{P}+\tilde{\rho}\right|^{2} \\
	&=\left|\mathcal{F}\left(P\right)\otimes\mathcal{F}\left(O_{\textrm{twin}}\right)+\tilde{\rho}_{\textrm{twin}}\right|^{2}, 
\end{align}
where $\otimes$ denotes convolution and we defined the twin object $O_\textrm{twin}=-O$ as well as the twin reference $\tilde{\rho}_\textrm{twin}=\tilde{P}+\tilde{\rho}$. $P$ and $\tilde{P}$ denote the probe and its Fourier transform, respectively. An analog argument holds for near-field diffraction geometries, where an additional quadratic phase envelope in the probe enters the math. It is thus seen that the twin object and the twin reference wave explain the same observed interferograms as the true object and reference. To avoid this ambiguity, a separate measurement of the reference wave (with the wave from the specimen blocked) can be carried out or a priori knowledge about the specimen can be provided (for example knowledge about the specimen being transparent in certain regions such as an empty microscopy slide).

\section{Scan grid optimization}
\label{sec: Scan grid optimization}

\begin{figure}[htbp]
  \centering
  \includegraphics[width=0.8\textwidth]{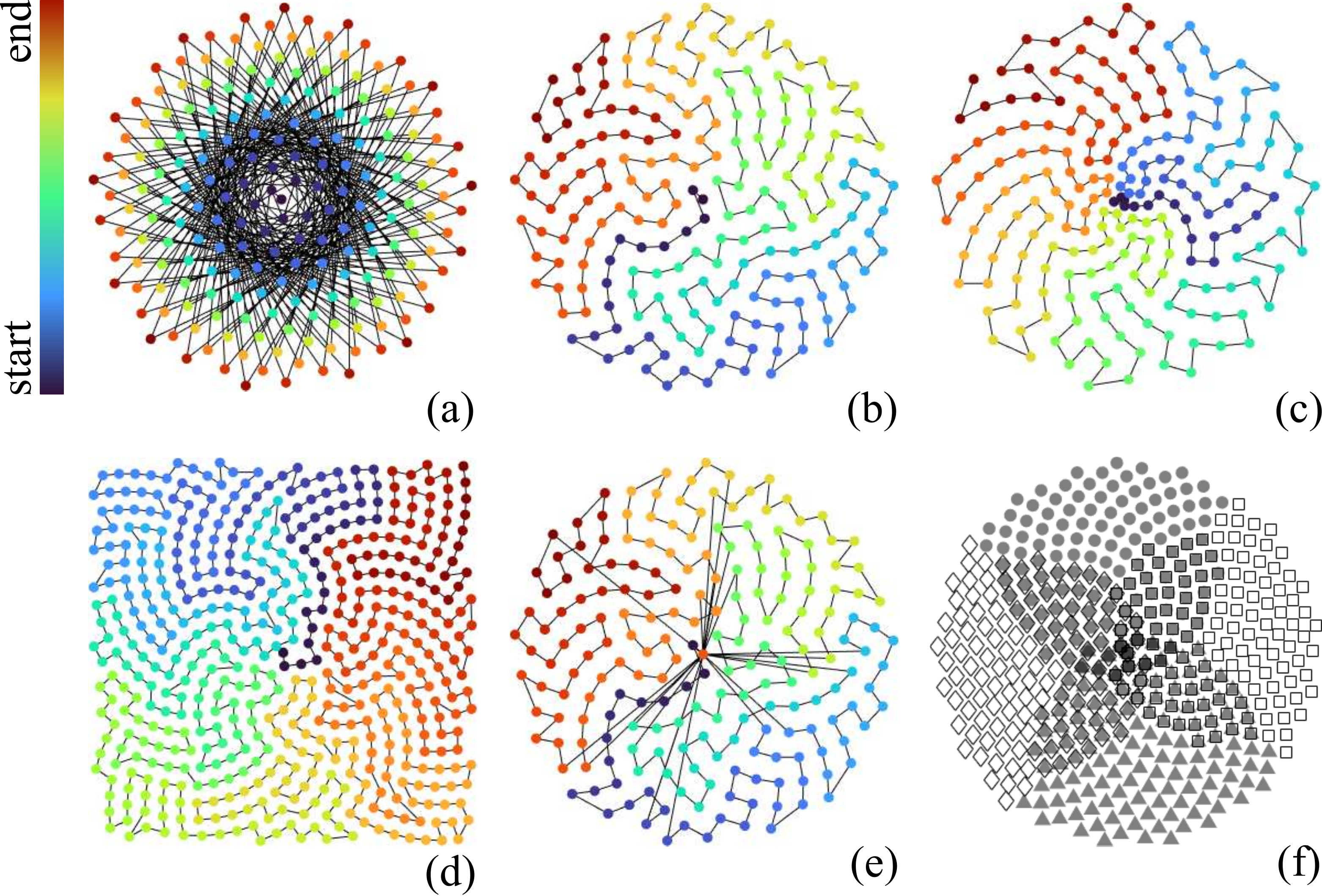}
\caption{ A variety of scan grids can be generated and optimized in \emph{PtyLab}. (a) The typical workflow is to generate an aperiodic scan grid in polar coordinates, here a Fermat grid, and subsequently preprocessing steps on it. (b) The total path of the scan trajectory is minimized by solving the traveling salesman problem. In some cases, morphological operations to scan grids are useful, such as non-uniform \emph{scaling} (c) and \emph{rectification} (d). (e) Another useful technique is \emph{checkpointing}, where the same scan point is revisited during a long scan. In panel (e) a Fermat grid with 200 scan points plus 20 checkpoints is shown, which are equally spaced in time. (f) For large scan grids an overlapping k-means (OKM) algorithm can be used to \emph{partition} the scan grid into overlapping clusters and subsequently process each cluster separately. The overlap between clusters is required to synchronize phase information, which can otherwise differ by a global offset, and for stitching a large-field-of-view image. }
  \label{fig: scan grids}
\end{figure}

In CP, a certain amount of consideration is needed to optimize the scan trajectory. To date, the majority of CP setups employ mechanical scanners, although variants exist where the beam is rapidly steered over the sample by means of galvo mirrors \cite{zhang2019field}. The latter offers advantages in terms of speed and overall cost of the experimental setup, but the isoplanatic illumination patch of such mirror systems is finite and thus limits the field of view over which the probe wavefront can be assumed to be stable, thus compromising one of the very benefits of CP. Hence mechanical scanners are still the preferred option. For such systems it is important to minimize the total scan distance in order to reduce scan time and prevent mechanical wear. In addition, unlike other scanning microscopy systems ptychography requires non-periodic scan grids to avoid ambiguities in the reconstruction \cite{Thibault2009_raster_grid}. A popular choice are Fermat scan grids \cite{Huang2014}, as depicted in Fig.~\ref{fig: scan grids}(a). This type of scan grid is conveniently described in polar coordinates, where its trajectory assumes the form of a spiral. Minimizing to total travel path can be done using a solver for the traveling salesman problem (TSP). In \emph{PtyLab}, we use a 2opt \cite{Croes1958} TSP heuristic solver, which offers a good compromise between optimality and optimization time. Fig.~\ref{fig: scan grids}(b) shows an example of a distance-optimized scan trajectory, where the color scale indicates the start (blue) and end position (red). 

Moreover, several operations are available to transform scan grids, including \emph{non-uniform scaling} and \emph{rectification} as shown Fig.~\ref{fig: scan grids}(c) and (d), respectively. The former allows for non-uniform spatial sampling, adjusted such that the sampling is higher in regions that are challenging to resolve, while the latter clips the field of view to a rectangular (here square) region. Another practically useful strategy is \emph{checkpointing} (see Fig.~\ref{fig: scan grids}(e)), which alters a given scan grid such that it revisits a certain reference point throughout the scan. Deviations in the diffraction data at the checkpoints allow for identifying sources of error in the experimental setup, including position drift, flux instability, and illumination wavefront variations. The checkpoints are equi-spaced in time. 

The aforementioned techniques are primarily scan grid preprocessing techniques, meaning that the scan grid is optimized prior to the actual experiment. After the data acquisition, scan grid postprocessing techniques may be required. For example, large scan grids can be \emph{partitioned} to prevent memory limitations. In this way large data sets may be chopped up into smaller pieces which are then individually processed \cite{Guizar-Sicairos2014}. It is important that the scan partitions spatially overlap, so that adjacent regions can be phase synchronized and stitched up post-reconstruction. To ensure overlap between the clusters, an overlapping k-means (OKM) algorithm can be used \cite{cleuziou2008extended,whang2015non}. Figure~\ref{fig: scan grids}(f) shows an example of a scan grid partitioned into four overlapping clusters (filled circles/triangles, unfilled diamonds/squares), each containing 160 scan points. In the middle, the clusters overlap. A second reason for scan grid partitioning can be to define batch gradients which speed up convergence and robustness \cite{Odstrcil2018}.

As a note on FP scan grids, at first sight it appears surprising that the technique does not exhibit raster scan artefacts although the most commonly employed LED arrays are typically regularly spaced. However, the typically regular LED spatial arrangement still corresponds to a non-periodic spacing in angle, which explains that it is not subject to the aforementioned raster scan artefacts. While most of the aforementioned preprocessing steps are not required for FP due to the absence of mechanical movement, checkpointing and partitioning may still be used for monitoring stability and distributed data analysis, respectively.

\section{Open experimental data and tutorials}
\label{sec: Open experimental data and tutorials}

\begin{figure}[htbp]
  \centering
  \includegraphics[width=\textwidth]{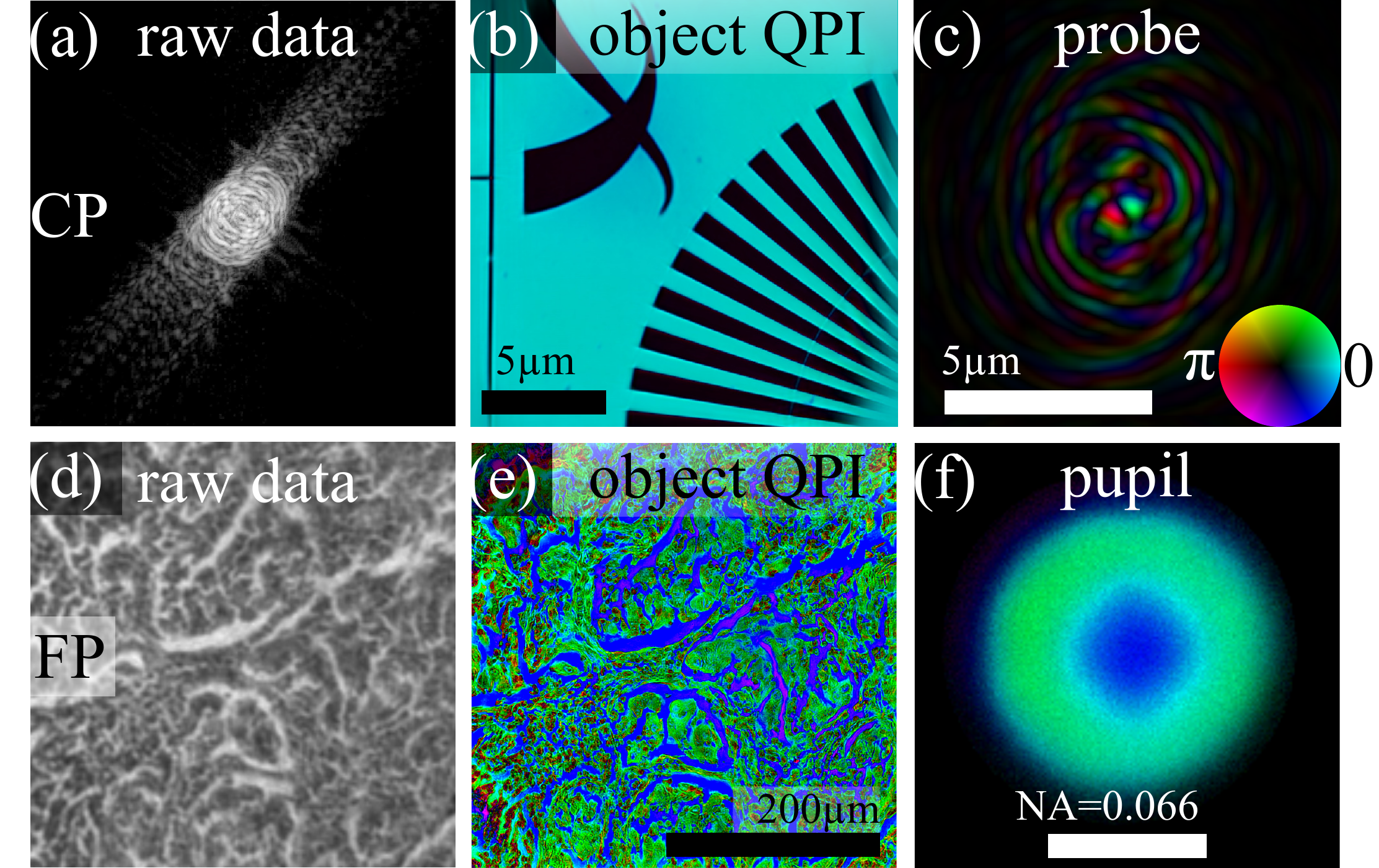}
\caption{ Examples of CP and FP experimental data analyses using \emph{PtyLab}. Top row: synchrotron-based soft x-ray CP (a) raw data, reconstructed (b) object QPI and (c) probe wavefront. Bottom row: visible light FP (d) raw data, reconstructed (e) object QPI and (f) pupil. Amplitude and phase are depicted as brightness and hue; experimental details in \cite{Loetgering2020helical, Aidukas2019multi}. Figure adapted from \cite{loetgering2021PtyLab}. }
\label{fig: experimental reconstructions}
\end{figure}

We publish a variety of CP and FP data sets and tutorials with the aim to introduce users to the functionality of \emph{PtyLab}. Figure \ref{fig: experimental reconstructions} depicts two such data sets. The top row shows a soft x-ray ($\lambda=2.48$nm) data set collected at a synchrotron (experimental details in \cite{Loetgering2020helical}). The bottom row depicts a visible light ($\lambda=625$nm) FP data set of lung carcinoma. For both data sets we show from left to right a single frame of the raw data, the recovered quantitative phase image (QPI) of the object (\emph{resolution test target} and \emph{lung carcinoma} histology slide), and reconstructed probe/pupil for the case of CP (top) and FP (bottom). Hue and brightness depict the phase and amplitude, respectively, of the complex-valued reconstructed quantities. A variety of additional data sets are published alongside \emph{PtyLab}, which are summarized in table \ref{tab: data overview}. Each of these data sets comes with an online tutorial explaining suitable data analysis approaches, including self-calibration and regularization.

\begin{table}[!t]
\begin{center}
\begin{tabular}{ |c|c|c|c|c| } 
 \hline
 mode & size (MB) & data set & tutorials & reference \\ 
 \hline
 CP & 360  & helical beam & regular reconstruction & \cite{Loetgering2020helical} \\ 
 CP & 360  & helical beam &  mixed-states & \cite{Loetgering2020helical} \\ 
 CP & 102  & USAF & axial position calibration (zPIE) & \cite{Loetgering2020zPIE}  \\ 
 CP & 404  & Siemens star & total-variation regularization & \cite{eschen2022material} \\ 
 FP & 10  & lung carcinoma & regular reconstruction & \cite{Aidukas2022addressing} \\ 
 FP & 10 & USAF & position calibration & \cite{Aidukas2022high} \\

 \hline
\end{tabular}
\end{center}
\caption{ Overview of open data sets and \emph{PtyLab} tutorials. 
}
\label{tab: data overview}
\end{table}

\section{Discussion and conclusions}

\emph{PtyLab} is a versatile ptychography software which we hope will aid researchers to explore the capabilities of CP and FP. Nevertheless, despite our excitement about this endeavor, we should mention some of its shortcomings: (1) Researchers with large-scale and high-throughput data analysis tasks (e.g. beamline scientists at synchrotrons) may be better off with one of the currently available high-performance ptychography packages mentioned in the introduction. However, we believe increased performance comes at the cost of flexibility in algorithm prototyping. (2) \emph{PtyLab} currently does not support tomographic reconstruction with specimen rotation. It is to be noted that external CT toolboxes, such as Astra \cite{van2016fast} or Tigre \cite{biguri2020arbitrarily}, can principally be used for ptychographic computed tomography once a sequence of 2D reconstructions at different angles is available. However, some ptychotomographic software embed specialized regularization techniques within the reconstruction routine \cite{odstrcil2019ab}, which are not available in standard CT packages. In contrast, ptychographic optical coherence tomography (POCT) \cite{Du2021} does not require angle diversity and 3D reconstructions can be obtained simply by performing a Fourier transform along the wavelength dimension in a multispectral reconstruction object stack. The latter is readily performed in \emph{PtyLab}.

We have designed \emph{PtyLab} based on the principle of reciprocity. An interesting implication of the conversion between CP and FP is performance, as it provides the freedom to choose whether the computational complexity of the Fourier transform operation, used in typical inversion algorithms, scales logarithmically with the number of pixels in the detector or with the number of scan positions in a given data cube - numbers which can be orders of magnitude apart so that even on a logarithmic scale, practical speed ups can be achieved. However, in order to take full advantage of reciprocity, interpolation techniques to non-equidistantly sampled geometries are required, which will be explored in future work.

In summary, we have presented \emph{PtyLab}, a cross-platform, open source inverse modeling toolbox for CP and FP. We believe \emph{PtyLab}'s major strengths lie in (1) the uniform framework for CP and FP enabling cross-pollination between the two domains, (2) the availability in three widely used programming languages (Matlab, Python, and Julia), making it easy for researchers with different programming backgrounds to exchange and benchmark code snippets and data anlyses, and (3) its versatile code architecture suited both for beginners and experts interested in rapid ptychographic algorithm prototyping. In addition, a plethora of self-calibration features (e.g. aPIE, zPIE) and algorithmic novelties (e.g. conversion between CP and FP, POCT, CP with external reference beam, sPIE) are available that to our knowledge have previously not been featured in open access ptychography code. Various functions for scan grid generation help the user to optimize data acquisition and postprocessing. For further information the reader is referred to the GitHub website with its accompanying tutorials as well as the open data provided along with it \cite{PtyLab}. 

\section*{Appendix}
\subsection*{Equivalence of CP and FP}
In this appendix, we provide a formal proof that the same data cube can be regarded as a CP or an FP data set, implying the ability to convert between the two.
Without loss of generality, we assume we are given a far-field CP data set
\begin{equation}
\label{eq: data cube}
    I_{d}\left(\boldsymbol{q},\boldsymbol{s}\right)=\left|\int P\left(\boldsymbol{x}\right)O\left(\boldsymbol{x}-\boldsymbol{s}\right)\exp\left[-i2\pi\boldsymbol{q}\boldsymbol{x}\right]d\boldsymbol{x}\right|^{2},
\end{equation}
where $\boldsymbol{s}$ and $\boldsymbol{q}$ denote scan positions and detector coordinates, respectively. For a given scan point $\boldsymbol{s}_0$, we have
\begin{align}
\label{eqn: CP forward}
    I_{d}\left(\boldsymbol{q},\boldsymbol{s}_{0}\right)	&=\left|\int P\left(\boldsymbol{x}\right)O\left(\boldsymbol{x}-\boldsymbol{s}_{0}\right)\exp\left[-i2\pi\boldsymbol{q}\boldsymbol{x}\right]d\boldsymbol{x}\right|^{2} \\
	&=\left|\tilde{P}\left(\boldsymbol{q}\right)\otimes\tilde{O}_{\boldsymbol{s}_{0}}\left(\boldsymbol{q}\right)\right|^{2},
\end{align}
where 
\begin{equation}
    \tilde{P}\left(\boldsymbol{q}\right)=\mathcal{F}_{\boldsymbol{x}\rightarrow\boldsymbol{q}}\left[P\left(\boldsymbol{x}\right)\right]
\end{equation}
is the probe spectrum and 
\begin{equation}
\tilde{O}_{\boldsymbol{s}_{0}}\left(\boldsymbol{q}\right)=\mathcal{F}_{\boldsymbol{x}\rightarrow\boldsymbol{q}}\left[O\left(\boldsymbol{x}-\boldsymbol{s}_{0}\right)\right]    
\end{equation}
is the object spectrum. CP solvers use the problem formulation in Eq.~\ref{eqn: CP forward} for a sequence of scan positions.

Next, consider a fixed observation pixel ($\boldsymbol{q}_{0}$) in the data cube in Eq.~\ref{eq: data cube}
\begin{align}
    I_{d}\left(\boldsymbol{q}_{0},\boldsymbol{s}\right)&=\left|\int P\left(\boldsymbol{x}\right)O\left(\boldsymbol{x}-\boldsymbol{s}\right)\exp\left[-i2\pi\boldsymbol{q}_{0}\boldsymbol{x}\right]d\boldsymbol{x}\right|^{2} \nonumber \\
    &=\left|\exp\left[-i2\pi\boldsymbol{q}_{0}\boldsymbol{s}\right]\int P\left(\boldsymbol{x}\right)O\left(\boldsymbol{x}-\boldsymbol{s}\right)\exp\left[-i2\pi\boldsymbol{q}_{0}\left(\boldsymbol{x}-\boldsymbol{s}\right)\right]d\boldsymbol{x}\right|^{2} \nonumber \\
    &=\left|\int P\left(\boldsymbol{x}\right)O\left(\boldsymbol{x}-\boldsymbol{s}\right)\exp\left[-i2\pi\boldsymbol{q}_{0}\left(\boldsymbol{x}-\boldsymbol{s}\right)\right]d\boldsymbol{x}\right|^{2} \nonumber \\
    &=\left|\int P\left(\boldsymbol{x}\right)O'_{\boldsymbol{q}_{0}}\left(\boldsymbol{s}-\boldsymbol{x}\right)d\boldsymbol{x}\right|^{2} \nonumber \\
    &=\left|P\left(\boldsymbol{s}\right)\otimes O'_{\boldsymbol{q}_{0}}\left(\boldsymbol{s}\right)\right|^{2}, \label{eq: FP forward}
\end{align}
where we defined
\begin{equation}
    O'_{\boldsymbol{q}_{0}}\left(\boldsymbol{x}\right)=O\left(-\boldsymbol{x}\right)\exp\left[i2\pi\boldsymbol{q}_{0}\boldsymbol{x}\right].
\end{equation}
FP solves the problem formulation in Eq.~\ref{eq: FP forward} for a sequence of illumination directions. Thus we may consider the same data cube to be either a CP or an FP inverse problem. From the CP perspective we reconstruct P and O, while from the FP perspective we reconstruct $\tilde{P}$ and $\tilde{O}$. Thus if we tackle a CP data from the FP perspective, we simply inverse Fourier transform the reconstructed object spectrum to retrieve the object that we would have reconstructed had we directly chosen a CP solver. A similar statement holds for the correspondence between probe and pupil.

\section*{Acknowledgements}

Python packages: numpy \cite{oliphant2006guide}, matplotlib \cite{hunter2007matplotlib}, h5py \cite{h5py}, scipy \cite{virtanen2020scipy}, scikit-image \cite{van2014scikit}, tqdm \cite{da2019tqdm}

\textit{PtyLab.jl} was implemented in Julia \cite{bezanson2017julia} and made use of the following packages: CUDA.jl\cite{Besard2018, Cuda.jl}, FFTW.jl \cite{frigo1998fftw, FFTW.jl} and LLVM.jl\cite{LLVM.jl}.

The Python, Matlab, and Julia versions of PtyLab are available online \cite{PtyLab}.

\section*{Funding Information}
SW acknowledges funding from the European Research Council (ERC-CoG 864016) and the Netherlands Organisation for Scientific Research NWO through the LINX Perspectief Programme. The work of TA is supported by funding from the Swiss National Science Foundation (SNF), Project Numbers $200021\_196898$

\section*{Disclosures}
The authors declare no conflicts of interest.


\bibliographystyle{unsrt} 
\bibliography{references}

\end{document}